\documentclass[12pt,preprint]{aastex}
\usepackage{bm}
\usepackage{threeparttable}
\usepackage{amsmath,amssymb}
\usepackage{textcomp}
\usepackage{booktabs}
\usepackage{lscape}
\usepackage{url}
\usepackage[usenames]{color}
\makeatletter
\newcommand{\tblcaption}[1]{\def\@captype{table}\caption{#1}}
\makeatother
\usepackage{subfigure}

\usepackage{datetime}

\newcommand\ltsima{$\; \buildrel <\over\sim \;$}
\newcommand\simlt{\lower.5ex\hbox{\ltsima}}
\newcommand\gtsima{$\; \buildrel >\over\sim \;$}
\newcommand\simgt{\lower.5ex\hbox{\gtsima}}
\newcommand\msun {M_\odot}

\begin{document}


\title{MOA-2016-BLG-227Lb: A Massive Planet Characterized by Combining Lightcurve Analysis and Keck AO Imaging}


\author{N. Koshimoto\altaffilmark{1,M}, Y. Shvartzvald\altaffilmark{2,K,W,U,a}, D. P. Bennett\altaffilmark{3,4,M}, M. T. Penny\altaffilmark{5,C,b},\\
M. Hundertmark\altaffilmark{6,7,V}, I. A. Bond\altaffilmark{8,M}, W. C. Zang\altaffilmark{9,10,C}, C. B. Henderson\altaffilmark{2,K,U,a},\\
D. Suzuki\altaffilmark{3,4,M}, N. J. Rattenbury\altaffilmark{11,M}, T. Sumi\altaffilmark{1,M}}

\and

\author{F. Abe\altaffilmark{12}, Y. Asakura\altaffilmark{12}, A. Bhattacharya\altaffilmark{3,4},  M. Donachie\altaffilmark{11}, P. Evans\altaffilmark{11},
A. Fukui\altaffilmark{13}, \\
Y. Hirao\altaffilmark{1}, Y. Itow\altaffilmark{12}, M.C.A. Li\altaffilmark{11}, C.H. Ling\altaffilmark{8}, 
K. Masuda\altaffilmark{12}, Y. Matsubara\altaffilmark{12}, T. Matsuo\altaffilmark{1}, \\
Y. Muraki\altaffilmark{12}, M. Nagakane\altaffilmark{1}, K. Ohnishi\altaffilmark{14}, C. Ranc\altaffilmark{3,a},
To. Saito\altaffilmark{15}, A. Sharan\altaffilmark{11},\\ 
H. Shibai\altaffilmark{1}, D.J. Sullivan\altaffilmark{16},  P.J. Tristram\altaffilmark{17},  T. Yamada\altaffilmark{1}, T. Yamada\altaffilmark{18}, A. Yonehara\altaffilmark{18}\\
(MOA Collaboration)}

\author{C. R. Gelino\altaffilmark{19,20},
C. Beichman\altaffilmark{19},
J.-P. Beaulieu\altaffilmark{21,22},
J.-B. Marquette\altaffilmark{21},
V. Batista\altaffilmark{21}\\
(Keck Team)}

\author{M. Friedmann\altaffilmark{23},
N. Hallakoun\altaffilmark{23,24},
S. Kaspi\altaffilmark{23},
D. Maoz\altaffilmark{23}\\
(Wise Group)}

\author{G. Bryden\altaffilmark{2},
S. Calchi Novati\altaffilmark{25,26},
S. B. Howell\altaffilmark{27}\\
(UKIRT Team)}

\author{T. S. Wang\altaffilmark{9},
S. Mao\altaffilmark{9,28,29},
P. Fouqu\'e\altaffilmark{30,31}\\
(CFHT-K2C9 Microlensing Survey)}

\author{H. Korhonen\altaffilmark{32}, 
U. G. J{\o}rgensen\altaffilmark{7}, R. Street\altaffilmark{33}, Y. Tsapras\altaffilmark{6,33}, M. Dominik\altaffilmark{34}, E. Kerins\altaffilmark{35},
A. Cassan\altaffilmark{36}, C. Snodgrass\altaffilmark{37}, E. Bachelet\altaffilmark{33}, 
V. Bozza\altaffilmark{26,38,39}, D. M. Bramich\altaffilmark{40}\\
(VST-K2C9 Team)}


\altaffiltext{1}{Department of Earth and Space Science, Graduate School of Science, Osaka University, 1-1 Machikaneyama, Toyonaka, Osaka 560-0043, Japan}
\altaffiltext{2}{Jet Propulsion Laboratory, California Institute of Technology, 4800 Oak Grove Drive, Pasadena, CA 91109, USA}
\altaffiltext{3}{Laboratory for Exoplanets and Stellar Astrophysics, NASA/Goddard Space Flight Center, Greenbelt, MD 20771, USA}
\altaffiltext{4}{Department of Physics, University of Notre Dame, Notre Dame, IN 46556, USA}
\altaffiltext{5}{Department of Astronomy, The Ohio State University, 140 W. 18th Avenue, Columbus, OH 43210, USA}
\altaffiltext{6}{Astronomisches Rechen-Institut, Zentrum f{\"u}r Astronomie der Universit{\"a}t Heidelberg (ZAH), 69120 Heidelberg, Germany}
\altaffiltext{7}{Niels Bohr Institute \& Centre for Star and Planet Formation, University of Copenhagen {\O}ster Voldgade 5, 1350 - Copenhagen, Denmark}
\altaffiltext{8}{Institute of Information and Mathematical Sciences, Massey University, Private Bag 102-904, North Shore Mail Centre, Auckland, New Zealand}
\altaffiltext{9}{Physics Department and Tsinghua Centre for Astrophysics, Tsinghua University, Beijing 100084, China}
\altaffiltext{10}{Department of Physics, Zhejiang University, Hangzhou, 310058, China}
\altaffiltext{11}{Department of Physics, University of Auckland, Private Bag 92019, Auckland, New Zealand}
\altaffiltext{12}{Institute for Space-Earth Environmental Research, Nagoya University, Nagoya 464-8601, Japan}
\altaffiltext{13}{Okayama Astrophysical Observatory, National Astronomical Observatory, 3037-5 Honjo, Kamogata, Asakuchi, Okayama 719-0232, Japan}
\altaffiltext{14}{Nagano National College of Technology, Nagano 381-8550, Japan}
\altaffiltext{15}{Tokyo Metropolitan College of Industrial Technology, Tokyo 116-8523, Japan}
\altaffiltext{16}{School of Chemical and Physical Sciences, Victoria University, Wellington, New Zealand}
\altaffiltext{17}{University of Canterbury Mt John Observatory, P.O. Box 56, Lake Tekapo 8770, New Zealand}
\altaffiltext{18}{Department of Physics, Faculty of Science, Kyoto Sangyo University, Kyoto 603-8555, Japan}
\altaffiltext{19}{NASA Exoplanet Science Institute, California Institute of Technology, Pasadena, CA 91125, USA}
\altaffiltext{20}{Infrared Processing and Analysis Center and NASA Exoplanet Science Institute, California Institute of Technology, Pasadena, CA 91125, USA}
\altaffiltext{21}{UPMC-CNRS, UMR 7095, Institut d’Astrophysique de Paris, 98Bis Boulevard Arago, F-75014 Paris, France}
\altaffiltext{22}{School of Physical Sciences, University of Tasmania, Private Bag 37 Hobart, Tasmania 7001 Australia}
\altaffiltext{23}{School of Physics and Astronomy and Wise Observatory, Tel-Aviv University, Tel-Aviv 69978, Israel}
\altaffiltext{24}{European Southern Observatory, Karl-Schwarzschild-Stra{\ss}e 2, D-85748 Garching, Germany}
\altaffiltext{25}{IPAC, Mail Code 100-22, Caltech, 1200 E. California Blvd., Pasadena, CA 91125}
\altaffiltext{26}{Dipartimento di Fisica ``E. R. Caianiello'', Universit\`a di Salerno, Via Giovanni Paolo II, 84084 Fisciano (SA),\ Italy}
\altaffiltext{27}{Kepler $\&$ K2 Missions, NASA Ames Research Center, PO Box 1,M/S 244-30, Moffett Field, CA 94035}
\altaffiltext{28}{National Astronomical Observatories, Chinese Academy of Sciences, A20 Datun Rd., Chaoyang District, Beijing 100012, China}
\altaffiltext{29}{Jodrell Bank Centre for Astrophysics, Alan Turing Building, University of Manchester, Manchester M13 9PL, UK}
\altaffiltext{30}{CFHT Corporation, 65-1238 Mamalahoa Hwy, Kamuela, Hawaii 96743, USA}
\altaffiltext{31}{Universit\'e de Toulouse, UPS-OMP, IRAP, Toulouse, France}
\altaffiltext{32}{Dark Cosmology Centre, Niels Bohr Institute, University of Copenhagen, Juliane Maries Vej 30, 2100 - Copenhagen {\O}, Denmark}
\altaffiltext{33}{Las Cumbres Observatory Global Telescope Network, Inc., 6740 Cortona Drive, Suite 102, Goleta, CA 93117, USA}
\altaffiltext{34}{SUPA, School of Physics \& Astronomy, University of St Andrews, North Haugh, St Andrews KY16 9SS, UK}
\altaffiltext{35}{Jodrell Bank Centre for Astrophysics, School of Physics and Astronomy, University of Manchester, Oxford Road, Manchester M13 9PL, UK}
\altaffiltext{36}{Sorbonne Universit{\'e}s, UPMC Univ Paris 6 et CNRS, UMR 7095, Institut
d’Astrophysique de Paris, 98 bis bd Arago, 75014 Paris, France}
\altaffiltext{37}{Planetary and Space Sciences, Department of Physical Sciences, The Open University, Milton Keynes, MK7 6AA, UK}
\altaffiltext{38}{Istituto Internazionale per gli Alti Studi Scientifici (IIASS), Via G.
Pellegrino 19, 84019 Vietri sul Mare (SA), Italy}
\altaffiltext{39}{INAF - Observatory of Capodimonte, Salita Moiariello, 16, 80131, Naples, Italy}
\altaffiltext{40}{Qatar Environment and Energy Research Institute(QEERI), HBKU, Qatar Foundation, Doha, Qatar}
\altaffiltext{M}{MOA Collaboration}
\altaffiltext{K}{Keck team}
\altaffiltext{W}{Wise group}
\altaffiltext{U}{UKIRT team}
\altaffiltext{C}{CFHT-K2C9 Microlensing Survey}
\altaffiltext{V}{VST-K2C9 Team}
\altaffiltext{a}{NASA Postdoctoral Program Fellow}
\altaffiltext{b}{Sagan Fellow}


\begin{abstract}
We report the discovery of a microlensing planet --- MOA-2016-BLG-227Lb --- with 
a  large planet/host mass ratio of $q \simeq 9 \times 10^{-3}$.
This event was  located near the {\it K2} Campaign 9 field
that was observed by a large number of telescopes.
 As a result, the event was in the microlensing survey area of a number of these telescopes,
and this enabled good coverage of the planetary light curve signal.
High angular resolution adaptive optics images from the Keck telescope reveal excess flux
at the position of the source above the flux of the source star, as indicated by the light 
curve model. This excess flux could be due to the lens star, but it could also be due to
a companion to the source or lens star, or even an unrelated star. We consider all these
possibilities in a Bayesian analysis in the context
of a standard Galactic model. Our analysis indicates that it is unlikely that a 
large fraction of the excess flux comes from the lens, unless solar type stars are
much more likely to host planets of this mass ratio than lower mass stars.
We recommend that a method similar to the one developed in this paper be used
for other events with high angular resolution follow-up observations when the follow-up
observations are insufficient to measure the lens-source relative proper motion.
\end{abstract}


\section{Introduction}
Gravitational microlensing is a powerful method for detecting extrasolar planets \citep{mao91,gouloe92,gau12}.
Compared to other detection techniques, microlensing is sensitive to low-mass planets \citep{ben96}
orbiting beyond the snow line around relatively faint host stars like M dwarfs or brown 
dwarfs \citep{ben08, sum16}, which is complementary to other methods.

A difficulty with the microlensing method is the determination of the mass of a lens $M_L$ 
and the distance to the lens system $D_L$.
If we have an estimate for the angular Einstein radius $\theta_E$ and the microlens 
parallax $\pi_E$, the mass is directly determined by 
\begin{equation}
M_L = \frac{\theta_{\rm E}}{\kappa \pi_{\rm E}} \ ,
\end{equation}
where $\kappa = 8.144 ~{\rm mas} ~M_\odot ^{-1}$ \citep{gou92,gaudi-ogle109,muraki11}. 
 When the source distance, $D_S \sim 8$ kpc, is known, the distance to the lens is given by
\begin{equation}
D_L = {{\rm AU} \over \pi_E \theta_E + \pi_S},
\end{equation}
where $\pi_S \equiv {\rm AU}/D_S$.
However the microlens parallax can be observed for a fraction of planetary events, 
while the angular Einstein radius is observed for most planetary events. 

One strategy  to estimate $M_L$ and $D_L$ for events in which microlens parallax cannot be 
detected is to  use a Bayesian analysis based on probability distributions from a standard
Galactic model (e.g., Beaulieu et al. 2006; Bennett et al. 2014; Koshimoto et al. 2014; 
Shvartzvald et al. 2014).  However, such an analysis must necessarily make an assumption
about the probability that stars of a given mass and distance will host a planet. The most
common assumption is that all stellar microlens stars are equally likely to host a planet
with the properties of the microlens planet in question. It may be that
the probability of hosting a planet of the measured mass ratio and separation depends on 
the host mass or the distance from the Galactic center. But, without mass and distance measurements,
these quantities are determined by our Bayesian prior assumptions.
As a case in point, \citet{ben14} analyzed MOA-2011-BLG-262 and found a planetary mass
host orbited by an Earth-mass ``moon" model had  almost the same likelihood as a star$+$planet 
model. But, since we have no precedent for such a rogue planet+moon system, they selected the
more conventional star$+$planet system as the favored model.
Also, the first discovered microlensing planet, OGLE-2003-BLG-235Lb, was at first thought to 
be a giant planet orbiting an M dwarf with a mass of $M_* \sim 0.36  M_{\odot}$ from a
Bayesian analysis \citep{bon04}. Such a system is predicted to be rare according to  
the core accretion theory of planet formation \citep{lau04, ken08}. 
 Follow-up {\it HST} images revealed a more massive host star
with mass of $M_* = 0.63^{+0.07}_{-0.09} M_{\odot}$ by  detecting 
excess flux in multiple passbands \citep{ben06}.

 When we measure the lens flux with high angular resolution {\it HST} or adaptive 
optics (AO) images (e.g., Bennett et al. 2006, 2007, 2015; Batista et al. 2015), we 
can then calculate the lens mass $M_L$ using  a mass-luminosity relation combined 
with the mass distance relation derived from $\theta_E$ measurement. High angular resolution
images are needed because microlensed source stars are generally located in dense Galactic bulge 
fields where there are usually multiple bright main sequence stars per ground-based 
seeing disk.

Because the size of the angular Einstein radius is $\simlt 1\,$mas, and the lens-source
relative proper motion is typically $\mu_{\rm rel} \sim 6\,$mas/yr,
it is possible that the lens and source stars will remain unresolved  even in high angular resolution 
images taken within a few years of the microlensing event. In such cases, there will be excess 
flux above that contributed by the source star and this excess flux must include the lens star flux. 
Some studies \citep{bat14,fuk15,kos17b}, which detected an excess flux, have assumed that 
this excess flux is dominated by the lens flux, and they have
derived the lens mass under this assumption. 

With this method, it  might seem that no assumptions are required regarding the probability
of the microlens stars to host planets, and there would be no biases due to any 
inadequacies of the Galactic model used.  However, \citet{bha17} use {\it HST} imaging
to show that the excess
flux at the position of the MOA-2008-BLG-310 source is {\it not} due to the lens star, and
\citet{kos17} have developed a Bayesian method to study the possibility of excess
flux from stars other than the lens star. Possibilities include unrelated stars, and companions 
to the source and lens stars. They find that it can be difficult to exclude all these contamination  
scenarios, especially for events with small angular Einstein radii.
In those cases where we cannot exclude the contamination scenarios, we can again use
a Bayesian analysis similar to the one described above to estimate the probability distribution 
of the lens properties. This means that we need to assume prior distributions for stellar binary
systems and the stellar luminosity function even when we have detected
excess flux in high angular resolution images. In cases where the lens properties are 
confirmed by a measurement of the lens-source relative proper motion \citep{ben15,bat15}
or microlensing parallax measurements \citep{gaudi-ogle109,beaulieu16,ben16}, this
contamination can be ruled out. Attempts at lens-source relative proper motion 
measurements can also confirm contamination \citep{bha17} in cases where the measured
proper motion of the star responsible for the excess flux does not match the microlensing
light curve prediction.

 In this paper, we report the discovery of the planetary microlensing event MOA-2016-BLG-227.
Observations and data reduction are described in Sections \ref{sec-obs} and \ref{sec-reduction}.
Our modeling results are presented in Section \ref{sec-model}.
In Section \ref{sec-thetaE}, we model the foreground extinction by comparing observed color
magnitude diagrams (CMDs) to different extinction laws and compare the results from the different 
extinction laws. Then, we use the favored extinction law to determine the 
angular Einstein radius, $\theta_{\rm E}$.
In Section \ref{sec-Keck}, we describe our Keck AO observations and photometry, 
and we  determine the excess flux at the position of the source.
In Section \ref{sec-lens}, we describe  our Bayesian method to determine the probability that
this excess flux is due to lens star and various combinations of other ``contaminating" stars. 
The posterior probabilities for this MOA-2016-BLG-227 planetary microlensing event are
presented, and we consider the effect of different planet hosting probability priors.
Finally, we discuss and conclude the results of our work in Section \ref{sec-disc}.

\section{Observations}\label{sec-obs}
The Microlensing Observations in Astrophysics (MOA; Bond et al. 2001, Sumi et al. 2003) 
group  conducts a high cadence survey towards the Galactic bulge using the 2.2-deg$^2$ FOV 
MOA-cam3 \citep{sak08} CCD camera mounted on the 1.8 m MOA-II telescope at the 
University of Canterbury Mt.\ John Observatory
 in New Zealand. The MOA group alerts about 600 microlensing events per year.
 Most observations are conducted in a customized MOA-Red filter which is similar to the sum 
of the standard Cousins $R$-and $I$-band filters. Observations with the MOA $V$ filter 
(Bessell $V$-band) are  taken once every clear night in each MOA field.

The microlensing event MOA-2016-BLG-227 was discovered  and announced 
by the MOA alert system \citep{bon01}
at (R.A., Dec.)$_{J2000}$ = (18:05:53.70, -27:42:51.43)  and $(l, b) = (3.303^{\circ},-3.240^{\circ})$ 
on 5 May 2016 (HJD$' \equiv $ HJD - 2450000 $\sim $ 7514).
This event occurred during the microlensing Campaign 9 of the {\it K2} Mission 
 ({\it K2}C9; Henderson et al. 2016)
and it was located close to (but not in) the area of sky that was surveyed for the
{\it K2}C9. This part of the {\it K2} field that was downloaded at
30 minute intervals is known as the ``superstamp\rlap." Because this event was so close 
to the superstamp, several other groups  conducting observing campaigns coordinated with
the {\it K2}C9 observations also observed this event.

The Wise group used the Jay Baum Rich   telescope, a Centurion 28
inch telescope (C28) at the Wise Observatory in Israel, which is equipped with a 1 deg$^2$ camera.  The group monitored the {\it K2}C9 superstamp during the campaign with six survey fields that  were observed 3--5 times per night with the Astrodon Exo- Planet BB (blue-blocking) filter. Although the MOA-2016-BLG-227 target was just outside the {\it K2}C9 superstamp, it was still within the Wise survey footprint.

The event was also observed with the wide-field near infrared (NIR) camera (WFCAM) on the 
UKIRT 3.8m telescope on Mauna Kea, Hawaii, as part of a NIR microlensing survey in 
conjunction with the {\it K2}C9 \citep{shv17} survey.
The UKIRT survey covered 6 deg$^2$, including the entire {\it K2}C9 superstamp and 
extending almost to the Galactic plane, with a cadence of 2--3 observations per night.
Observations were taken in $H$-band, with each epoch composed of sixteen 5-second 
co-added dithered exposures (2 co-adds, 2 jitter points, and $2\times2$ microsteps).

The Canada France Hawaii Telescope (CFHT), also on Mauna Kea, 
serendipitously  observed the event during the CFHT-{\it K2}C9 Microlensing Survey. The CFHT operated 
a multi-color survey of the  {\it K2}C9 superstamp using the Megacam Instrument \citep{bou03}. 
The CFHT observations for the event were conducted through the $g$-, $r$- and $i$-band filters.

 The VLT Survey Telescope (VST) is a 2.61m telescope installed at ESO's 
Paranal Observatory, and it carried out {\it K2}C9 observations as a 99-hours filler 
program \citep{arn98, kui02}. Observations for such a filler program could only be carried out
whenever the seeing was worse than 1 arcsec or conditions were
non-photometric. The main objective of the microlensing program was to monitor 
the {\it K2}C9  superstamp in an automatized mode to improve the event coverage and to 
secure color-information in SDSS $r$ and Johnson $V$ passbands.
Due to weather
conditions, Johnson $V$ images were only taken in the second half of the {\it K2}C9 survey,
and therefore MOA-2016-BLG-227 is only covered by SDSS $r$. The exact
pointing strategy was adjusted to cover the superstamp with 6 pointings
and to contain as many microlensing events from earlier seasons as
possible. In addition, a two-point dither  was obtained to reduce the
impact of bad pixels and detector gaps. Consequently, some events, like
MOA-2016-BLG-227, received more coverage and have been observed with
different CCDs.

\begin{figure}
\centering
\epsscale{0.7}
\rotatebox{-90}{
\plotone{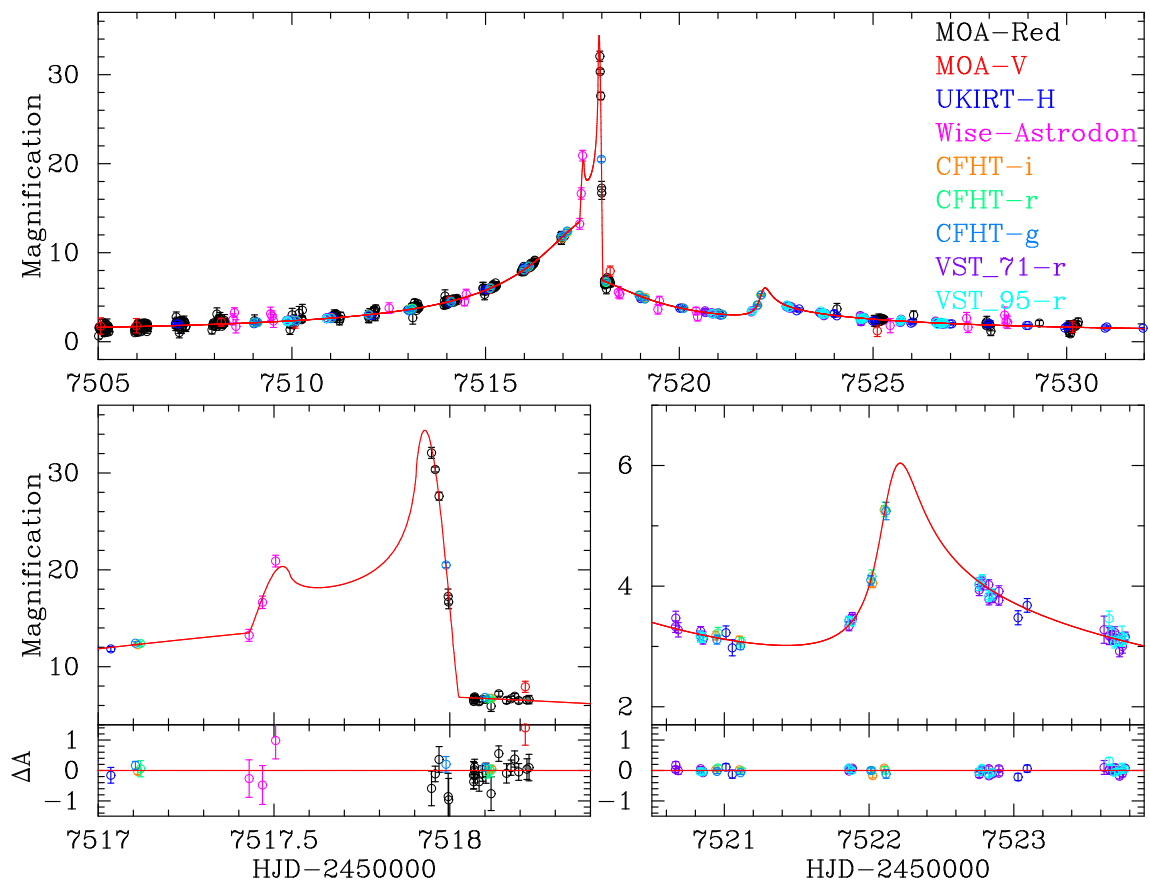}
}
\caption{The light curve data for MOA-2016-BLG-227 is plotted with the best-fit model.
The top panel shows the whole event, the bottom left and bottom right panels highlight the 
caustic crossing feature and the second bump due to the cusp approach, respectively.
The residuals from the model are shown in the bottom insets of the bottom panels.}
\label{fig-models}
\end{figure}

Figure \ref{fig-models} shows the observed MOA-2016-BLG-227 light curve.
MOA   announced the detection of a light curve anomaly for this 
event on 9 May 2016   (HJD$' = $ HJD $-2450000 \sim $ 7518),
and identified the anomaly as a planetary signal 4.5 hours after the anomaly alert.
 Although MOA detected a strong planetary caustic exit, the observing conditions were poor 
at the MOA observing site both immediately before and after this strong light curve feature.
Fortunately, the additional observations
from the Wise, UKIRT, CFHT and VST telescopes  covered 
the other important features of the light curve.

\section{Data Reduction} \label{sec-reduction}

Photometry of the MOA,  Wise and  UKIRT data were conducted using the offline 
difference image analysis pipeline of  \citet{bon17} in which stellar images are 
measured using an analytical PSF model of the form used in the DoPHOT photometry code \citep{sch93}.

Differential flux lightcurves of the CFHT data were produced from Elixir 
calibrated images\footnote{\url{http://www.cfht.hawaii.edu/Instruments/Elixir/}} 
using a custom difference imaging analysis pipeline based on ISIS version 2.2 \citep{ala98, ala00} and utilizing 
an improved interpolation routine\footnote{\url{http://verdis.phy.vanderbilt.edu/}} \citep{siv12, ber96}. 
Further details of the CFHT data reduction will be presented in a future paper.

Since there is no public VST instrument pipeline, calibration images
from ESO's archive were used and combined.
 Restrictive bad pixel masks were extracted to prevent inclusion of flatfield 
pixels with $> 1\,\%$ nightly variation or with  $>10\,\%$ deviation from the average.
The calibrated images were reduced with the difference imaging
package DanDIA \citep{bra08}, which uses a numerical kernel for difference imaging and 
the routines from the RoboNet pipeline for photometry\citep{tsa09}.


It is known that error bars estimated  by crowded field photometry codes can be under or
overestimated depending on the specific details of event. The error bars
provided by the photometry codes are sufficient to find the best fit models, but
they do not allow a proper determination of the microlensing light curve model
parameter uncertainties. Therefore, we empirically normalize the   error bars for each data set. 
We used the formula presented in \citet{yee12} for normalization, 
$\sigma _i' = k \sqrt{\sigma^2_i + e_{\rm min}^2}$
where $\sigma_i$ is the original error of the $i$th data point in magnitudes, and the parameters 
for normalization are $k$ and $e_{\rm min}$.   The parameters
$k$ and $e_{\rm min}$ are adjusted so that the cumulative 
$\chi^2$ distribution as a function of the number of data points sorted 
by each magnification of the preliminary best-fit model is a straight line of slope 1. 

  The dataset used for our analysis and the obtained normalization parameters are 
summarized in Table \ref{tab-data}.

\section{Modeling} \label{sec-model}
 The modeling of a binary-lens event requires following parameters: the time of the source 
closest approach to the 
lens center of mass, $t_0$, the impact parameter, $u_0$, of the source trajectory with respect 
to the center of mass of the lens system, the Einstein radius crossing time 
$t_{\rm E}=\theta_{\rm E}/\mu_{\rm rel}$, the lens mass ratio,
$q \equiv M_{\rm p}/M_{\rm host}$ the separation of the lens masses, $s$, the angle between
the trajectory and the binary lens axis, $\alpha$, and the source size $\rho \equiv \theta_*/\theta_{\rm E}$.
The parameters $u_0$, $s$, and $\rho$ are given in units of the Einstein radius, and 
$M_{\rm host}$ and $M_{\rm p}$ are the masses of the host star and its planetary companion.
With these seven parameters, we can calculate the magnification as a function of time $A(t)$. 
In the crowded stellar fields where most microlensing events are found, most source stars are blended
with one or more other stars, so that we cannot determine the source star brightness directly
from images where the source is not magnified.  Therefore, we add another set of linear
parameters for each data set, the source and blend fluxes, $f_S$ and $f_b$,  which are related to the observed flux by $F(t) = f_S A(t) + f_b$.

When we include the finite source effect, we must consider limb darkening effects.
We adopt a linear limb-darkening law with one parameter, $u_{\lambda}$, for each data set.
 From the intrinsic source color, discussed in Section \ref{sec-color}, we choose the  
atmospheric parameters for stars with similar intrinsic color from \citet{ben13}. This yields
an effective temperature of $T_{\rm eff} \sim 5500$ K, 
 a surface gravity of $\log~ [g/({\rm cm~ s^{-2}})]= 4.0$, a metallicity of [M/H] = 0.0, and 
a microturbulence velocity of $\xi = 1.0~ {\rm km ~s^{-1}}$.
We select the limb-darkening coefficients from the ATLAS model by \citet{cla11}  using these atmospheric parameters.
We have
$u_{\rm MOA-Red} = 0.5585$ for MOA-Red, $u_V = 0.6822$ for MOA-$V$, $u_R = 0.6015$ for 
Wise Astrodon, $u_H = 0.3170$ for UKIRT $H$, 
$u_i = 0.5360$ for CFHT $i$, $u_r = 0.6257$ for CFHT $r$, VST-71 $r$ and VST-95 $r$, 
and $u_g = 0.7565$ for CFHT $g$.
We used the mean of the $u_I$ and $u_R$ values for the limb-darkening coefficients 
for the MOA-Red passband.
Here we adopted the $R$-band limb-darkening coefficient for the Wise Astrodon data. 
As the Wise Astrodon filter is non-standard, our choice is not perfect. 
However we note that even if we adopt $u = 0$ for the limb darkening coefficient used 
with the Wise data, our $\chi^2$ value changes by only 1.5.
 The limb-darkening coefficients are also listed in Table \ref{tab-data}.

\begin{figure}
\centering
\epsscale{0.6}
\rotatebox{-90}{
\plotone{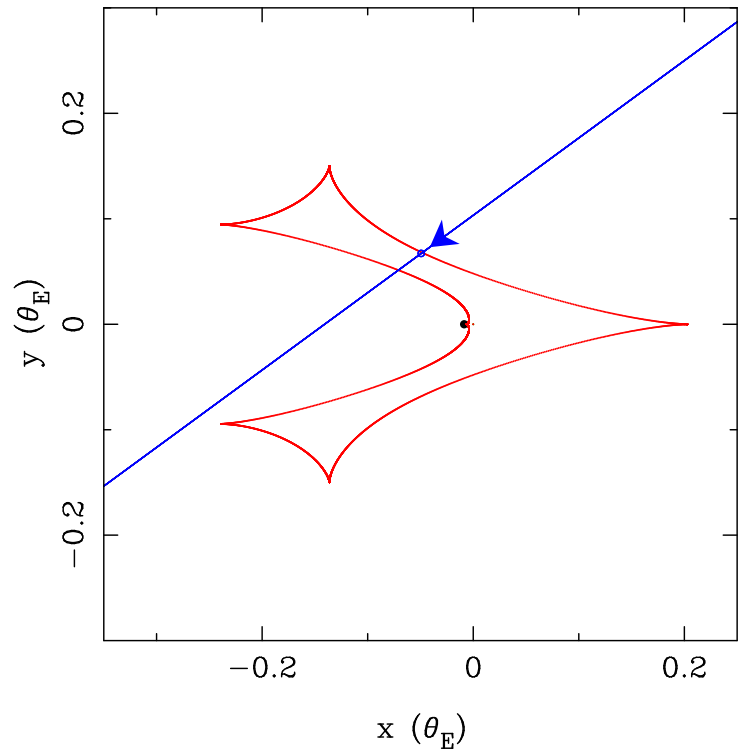}
}
\caption{The caustic curve for the best-fit model. 
The blue arrowed line indicates the source trajectory and the tiny 
blue circle on the caustic entry indicates the source size.}
\label{fig-caus}
\end{figure}

To find the best-fit model,  we conduct a global grid search using the method of 
\citet{sum16} where we fit the light curves using the Metropolis algorithm \citep{metrop}, 
with magnification calculations from the image centered ray-shooting method \citep{ben96,ben10}.
 From this, we find a unique model in which the source crosses the resonant caustic.
We show the model light curve in Figure \ref{fig-models}, the caustic and the source trajectory 
in Figure \ref{fig-caus} and the best-fit model parameters in Table \ref{tab-models}  along
with the parameter error bars, which are calculated with
a Markov Chain Monte Carlo (MCMC) \citep{ver03}.

We  also model the light curve including the microlensing parallax effect due
to the Earth's orbital motion \citep{gou92,alc95} although this event is unlikely to reveal 
a significant microlensing parallax signal because of its relatively short timescale.
 We find that the inclusion of the parallax effect improves the fit by $\Delta \chi^2 \sim 14$.
However, the parts of the lightcurve which contribute to this decrease in $\chi^2$ have 
a scatter similar to the variability of the  MOA baseline data,
and the best fit microlensing parallax parameter  is abnormally
large, $\pi_E = 1.3^{+2.1}_{-0.3}$, yielding a very small lens mass of $M_L \sim 0.02 M_{\odot}$.
Therefore, we conclude that the improvement of the fit by the parallax effect is due to systematic
errors in the MOA baseline data.

\section{Angular Einstein Radius} \label{sec-thetaE}

Because we have measured the finite source size, $\rho$, to a precision of $\sim 2\,$\%, the
determination of the angular source star radius $\theta_*$ will yield the angular
Einstein radius $\theta_E = \theta_*/\rho$. This, in turn, provides the mass-distance relation,
\citep{bennett_rev,gau12}
\begin{equation}
M_L = {c^2\over 4G} \theta_E^2 {D_S D_L\over D_S - D_L} 
       = 0.9823\,\msun \left({\theta_E\over 1\,{\rm mas}}\right)^2\left({x\over 1-x}\right)
       \left({D_S\over 8\,{\rm kpc}}\right) \ ,
\label{eq-m_thetaE}
\end{equation}
where $x = D_L/D_S$.
We can empirically derive $\theta_*$ from the intrinsic source magnitude and the color
\citep{kervella_dwarf,boy14}.

\subsection{Calibration} \label{sec-color}
Our fist step is to calibrate the source magnitude to a standard photometric system.
We cross referenced stars in the event field between our DoPHOT photometry catalog of 
stars in the MOA image and the OGLE-III catalog \citep{szy11} to convert MOA-Red and MOA $V$ 
into standard magnitudes. 
Following the procedure presented  in \citet{bon17}, we find the relations
\begin{align}
I_{\rm OGLE-III} - R_{\rm MOA} &= (28.186 \pm 0.006) - (0.247 \pm 0.005) (V - R)_{\rm MOA} \\
V_{\rm OGLE-III} - V_{\rm MOA} &= (28.391 \pm 0.004) - (0.123 \pm 0.004)  (V - R)_{\rm MOA}.
\end{align}
Using these calibration formulae and the result of light curve modeling, we  obtain 
the source star magnitude $I_S = 19.536 \pm 0.019$ and the color $(V-I)_S = 1.60 \pm 0.03$.

We follow a similar procedure to 
cross referenced stars in our DoPHOT photometry catalog of stars in the UKIRT images to stars 
in the VVV \citep{min10} catalog which is calibrated to the Two
Micron All Sky Survey (2MASS) photometric system \citep{car01}, thereby obtaining the 
relationship between these photometric systems. We use this same VVV catalog to plot CMDs 
in the next section and for the analysis of the Keck images in Section \ref{sec-Keck}.
Using the UKIRT $H$-band source magnitude obtained from the light curve model and 
the calibration relation, we find $H_S = 17.806 \pm 0.017$.
We also measure the colors of the source star: $(V-H)_S = 3.33 \pm 0.03$ and $(I-H)_S = 1.730 \pm 0.017$.

\subsection{Extinction and the angular Einstein radius} \label{sec-ext}
Next, we correct for extinction following the standard procedure \citep{yoo04,bennett10} 
 using the centroid of red giant clump (RGC) in the CMD as a standard candle.

\subsubsection{ RGC centroid measurement}
\begin{figure}
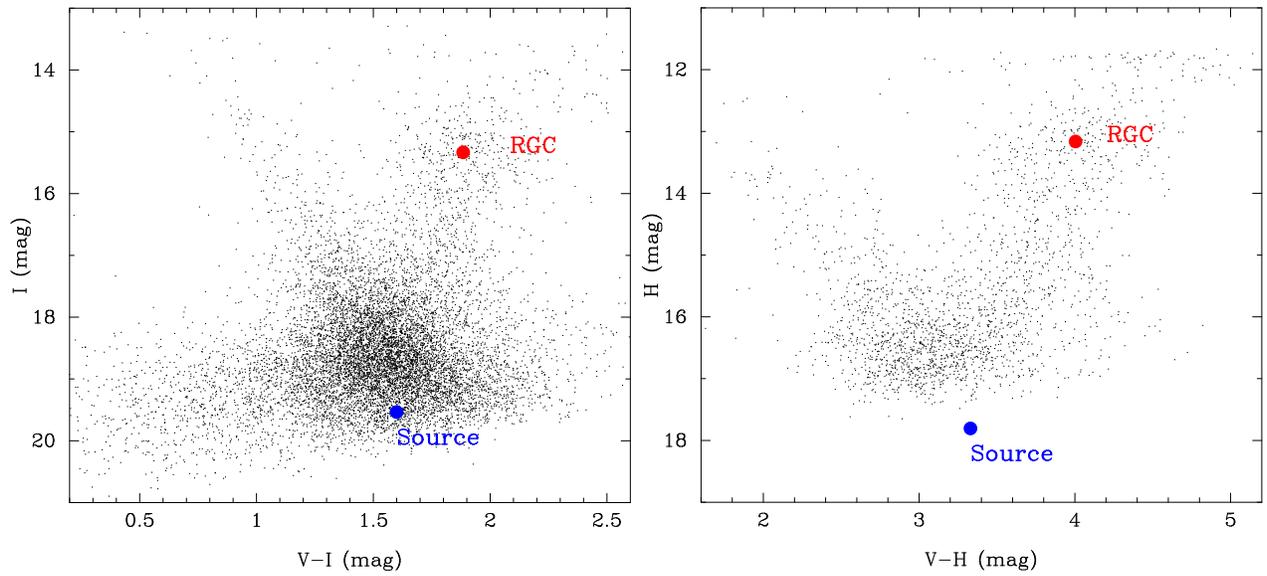

\begin{minipage}{0.5\hsize}
\begin{center}
\rotatebox{-90}{
\includegraphics[width=75mm]{cmd_VI.eps}
}
\end{center}
\end{minipage}
\begin{minipage}{0.49\hsize}
\begin{center}
\rotatebox{-90}{
\includegraphics[width=75mm]{cmd_VH.eps}
}
\end{center}
\end{minipage}
\caption{The color magnitude diagrams (CMDs) of stars within $2^{\prime}$ of the source star.
The left panel shows $V-I$ vs $I$ for the stars in OGLE-III catalog  \citep{szy11}, and the right panel
shows $V-H$ vs $H$ using stars from the  OGLE-III catalog to the VVV catalog,
which is calibrated to the 2MASS magnitude scale.
The source star and the mean of red giant clump are shown as the blue and red dots, respectively.}
\label{fig-cmd}
\end{figure}

Figure \ref{fig-cmd} shows the ($V-I$,$I$) and ($V-H$,$H$)  CMDs for stars within 
2 arcmin of the source star. The $V$ and $I$ magnitudes are taken from
the OGLE-III photometry catalog \citep{szy11}, and the VVV \citep{min10} catalog to the 
2MASS photometry scale for $H$-band magnitudes.
To plot the $V-H$ vs $H$ CMD, we cross referenced stars in the VVV catalog to stars in the OGLE-III catalog.
 For this cross reference, we use only isolated stars that are cross-matched to within 
1 arcsec of stars in the OGLE-III catalog to  ensure one-to-one matching between the two
catalogs. We note that the 1-arcsec limits corresponds to the average seeing in the VVV images.
We find the centroids of RGC in the $(V-I, I)$ and $(V-H, H)$ CMDs are $I_{\rm cl} = 15.33 \pm 0.05$,
$(V-I)_{\rm cl} = 1.88 \pm 0.02$,  $(V-H)_{\rm cl} = 4.03 \pm 0.06$ and $(I-H)_{\rm cl} = 2.11 \pm 0.03$.

\subsubsection{ RGC intrinsic magnitude and color}
We use $(V - I)_{\rm cl, 0} = 1.06 \pm 0.03$ and $I_{\rm cl, 0} = 14.36 \pm 0.05$ for the intrinsic 
$V-I$ color and $I$ magnitude of the  RGC \citep{ben13,nat16} at the Galactic longitude
of this event.
Following \citet{nat16}, we calculate the intrinsic color of $V-H$ and $I-H$ in the photometric system 
we are using now (i.e., Johnson $V$, Cousins $I$ and 2MASS $H$) by the tool provided by 
\citet{cas14} which is based on a grid of MARCS model atmospheres \citep{gus08}.
Assuming the stellar atmospheric parameters [Fe/H] = $-0.07 \pm 0.10$ \citep{gon13},  
$\log~ g = 2.3 \pm 0.1$ and 
[$\alpha$/Fe] = $0.20 \pm 0.05$ \citep{hil11, joh14} for the  RGC in the event field,
we derive $(V-H)_{\rm cl, 0} = 2.36 \pm 0.09$ and $(I-H)_{\rm cl, 0} = 1.30 \pm 0.06$ by adjusting
 the last atmospheric parameter $T_{\rm eff}$ so that 
the  $(V-I)$ value is in  the range of $1.03 < (V - I) < 1.09$.
We summarize the magnitude and colors for the RGC centroid
and the source in Table \ref{tab-RCG}.

\subsubsection{Angular Einstein radius}
By subtracting the intrinsic RGC color and magnitude values from the measured 
RGC positions in our CMDs, we  find an extinction values of
$A_{I, {\rm obs}} = 0.98 \pm 0.07$, and color excess values of 
$E(V-I)_{\rm obs} = 0.82 \pm 0.04$, $E(V-H)_{\rm obs} = 1.67 \pm 0.11$ and 
$E(I-H)_{\rm obs} =0.81 \pm 0.07$.
Following the method of \citet{bennett10}, we fit these values to the extinction laws 
of \citet{car89}, \citet{nis09} and \citet{nis08} separately and compared the results.
We present this analysis in Appendix \ref{sec-complaw}.
 From this comparison of models, we choose the \citet{nis08} extinction law, which yields 
an H-band extinction of $A_H = 0.19 \pm 0.02$ and a source angular radius of
$\theta_* = 0.68 \pm 0.02 ~ \mu$as. This $\theta_*$ value implies an
angular Einstein radius of $\theta_E = \theta_*/\rho = 0.227^{+0.006}_{-0.009}$ mas and a
lens-source relative proper motion of
$\mu_{\rm rel} = \theta_E / t_E = 4.88^{+0.14}_{-0.17}$ mas/yr.

\section{Excess Flux from Keck AO Images} \label{sec-Keck}

On August 13, 2016 (HJD$'$ = 7613.85) we observed MOA-2016-BLG-227 using 
the NIRC2 camera and the laser guide star (LGS) adaptive optics (AO) system mounted 
on the Keck II telescope at Mauna Kea, Hawaii.
Observations were  conducted in the $H$-band using the wide-field camera (0.04''/pix).
We took four dithered frames with 5 sec exposures and three additional dithered 
frames with a total integration time of 90 sec (6 co-adds of 15 sec exposures).
The first set of these images allows photometric calibration using unsaturated
bright stars, and the second set provides the increased photometric sensitivity to 
provide a high signal-to-noise flux measurement of the target.
Standard dark and flat field corrections were applied to the images, and sky subtraction was done 
using a stacked image from a nearby empty field.
Each set of images was then astrometrically aligned and stacked.
Finally, we use SExtracor \citep{ber96} to extract the Keck source catalog from the stacked images.

\begin{figure}
\centering
\epsscale{0.7}
\rotatebox{-90}{
\plotone{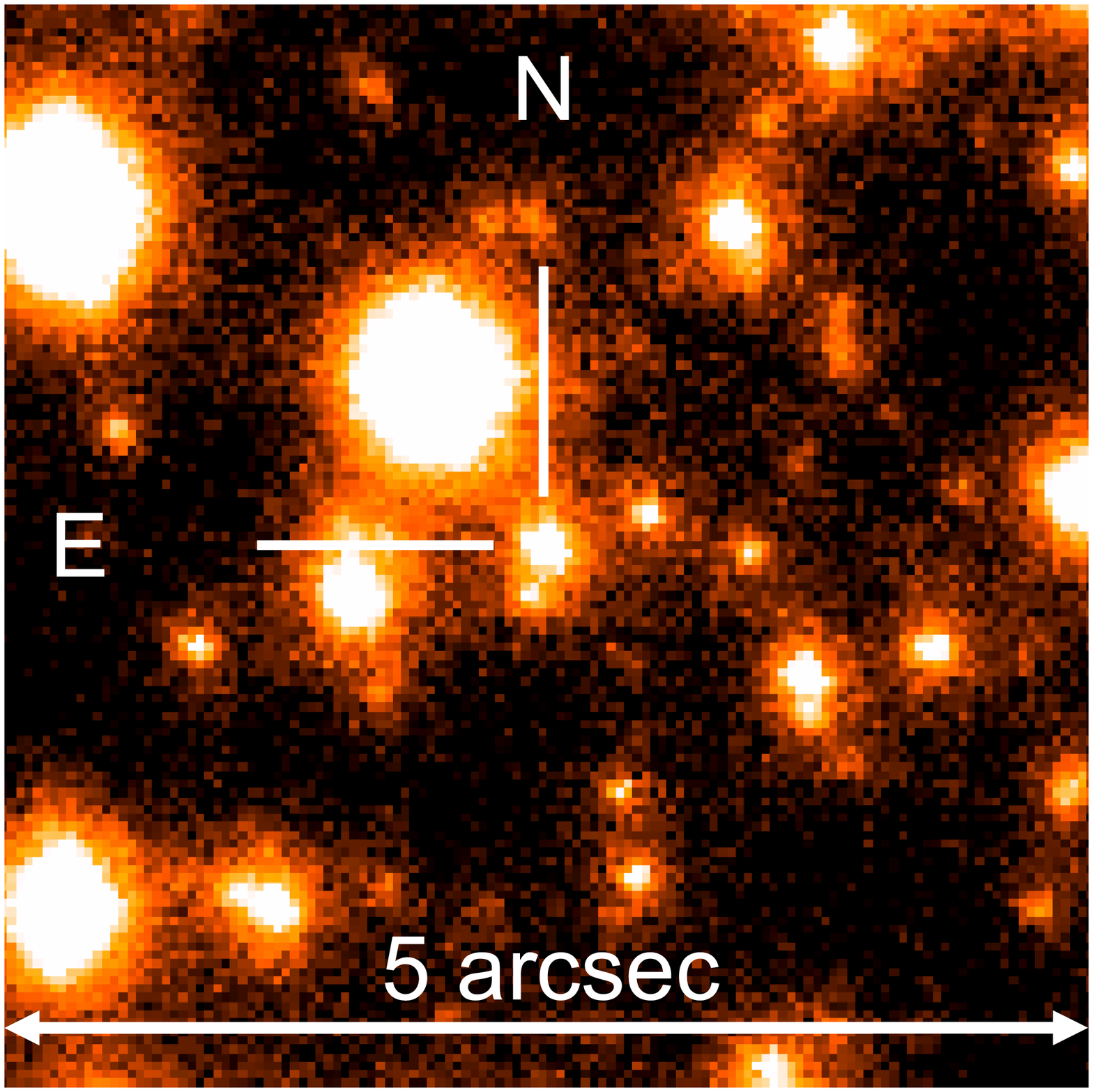}
}
\caption{The co-added Keck II AO image of the event field. The target is indicated.}
\label{fig-Keck}
\end{figure}

A calibration catalog was extracted using an $H$-band image of the target area taken 
by the VISTA Variables in the Via Lactea survey (VVV; Minniti et al. 2010) reprocessed following the
approach described in \citet{beaulieu16}.
We apply a zero point correction for the Keck source catalog using common VVV and Keck sources.
The estimated  zero point uncertainty is 0.05.  Figure \ref{fig-Keck} shows the Keck II AO image of the 
field. It indicates a bright star close to the target. As a result, the dominant photometry error
 comes from the background flux in the wings of the PSF of the nearby star.

We  determine the source coordinates  from a MOA difference image of the event 
while it was highly magnified.
We then identify the position of the microlensing target (source+lens) on the Keck image 
(see Figure \ref{fig-Keck}). The measured brightness of the target is $H_{\rm Keck}=17.63\pm 0.06$.
Due to technical problems in the AO system, the stellar images display sparse halo around each object.
Thus, the FWHM of the Keck image is 0.184$''$ (measured as the average of isolated bright 
stars near the target). This sets a limit on our ability to exclude flux contribution from stars 
unrelated to the source and the lens, as we discuss below.

 The light curve analysis of the UKIRT data $H$-band data  implies an (extinction uncorrected) 
$H$-band source magnitude of $H_S = 17.806 \pm 0.017$ (see Section \ref{sec-color}).
Because the Keck  observations were taken after the event reached its baseline brightness
($t_{\rm obs, Keck} - t_0 = 5.7~t_E$), we can extract the excess flux by subtracting 
the source flux from the target flux.  That is, 
$H_{\rm ex,obs} = H_S - 2.5 \log (F_{\rm Keck}/F_S - 1) = 19.7 \pm 0.4$, 
where $F_{\rm Keck}/F_S = 10^{-0.4(H_{\rm Keck} - H_S)}$.

\section{Lens Properties through Bayesian Analysis} \label{sec-lens}
\citet{kos17} present a systematic Bayesian analysis for the identification of the
star or stars producing excess flux at the position of the source seen in high-angular resolution images.
This analysis gives us the posterior probability distributions for the lens mass and the distance by
combining the results of the light curve modeling and the measured excess flux value.
The method is summarized as follows. 
\begin{enumerate}
\item  Determine prior probability distributions for four possibilities for the origin of the 
excess flux: the lens star, unrelated ambient stars, source companions or lens companions.
We denote these fluxes
by $F_L$, $F_{\rm amb}$, $F_{SC}$ and $F_{LC}$, respectively.
\item  Determine all combinations of the flux values for each type of star in the
prior distribution that are consistent with the observed excess flux,
$F_{\rm excess} = F_L + F_{\rm amb} + F_{SC} + F_{LC}$.
\end{enumerate}
The extracted combinations at step 2 corresponds to the posterior probability distributions 
for the MOA-2016-BLG-227 event.

\subsection{Prior probability distributions}
Now, we must determine the prior probability distributions of the four types of stars that 
can contribute to the excess flux at the position of the source.
We use all the information we have about this event --- except for the value of excess flux --- to 
create our prior probability distributions. This means that we include the FWHM of the Keck images, but not the
measured magnitude of the object at the location of the microlensing event.
Table \ref{tab-assumption} shows a summary of our assumptions.

\subsubsection{Lens flux prior}
For the lens flux prior distribution, we conduct a Bayesian analysis using the observed $t_E$ and 
$\theta_E$ values and the Galactic model, which has been used in a number of previous papers 
\citep{alc95,bea06} to estimate lens properties  for events with no microlensing parallax signal.
We use the Galactic model of \citet{han95} for the density and the velocity models and use the mass
function presented in the Supplementary Information section of \citet{sum11}.
Using this result and the mass-luminosity relation presented in \citet{kos17}, we  obtain the prior 
distribution for the lens apparent magnitude, $H_L$.
We adopt the formula for the extinction to the lens, 
$A_{H,L} = (1- e^{-D_L/h_{\rm dust}})/(1- e^{-D_S/h_{\rm dust}})~ A_{H,S}$, following \citet{ben15}, 
where $h_{\rm dust} = (0.1~ {\rm kpc})/\sin{|b|}$ is  the scale length of the dust toward the Galactic bulge,
 assuming a scale height of $0.1\,$kpc.
Note that this Bayesian analysis gives us  prior distributions for $M_L$, $D_L$ and $D_S$, in
addition to the $H_L$ prior distribution, but based on the observed $t_E$ and $\theta_E$ values. 
These values are needed for the calculation of the probability distributions below.

\subsubsection{Ambient star flux prior}
In order to determine the prior probability distribution for the flux of any unrelated ambient stars, we 
determine the number density  of stars in Keck AO images, centered on the target, within a magnitude range 
selected to have high completeness and divide that number by the area of the image. Then we use the
luminosity function of \citet{zoc03} to derive the number density of stars as a function of $H$ magnitude,
normalized to this measured number density in the Keck AO image.
 In this calculation, we correct for the differences in extinction and distance moduli between 
our field and that of \citet{zoc03},
using the distance moduli from Table 3 of \citet{nat13} and extinction values for both fields.

When correcting for the extinction difference, we also consider the difference between the extinction 
laws used.  \citet{zoc03} derived an $A_H$ value using the C89 extinction law 
with $R_V = 3.1$, whereas our preferred N08 extinction law implies a significantly different $A_H$ value.
To correct for this difference, we calculate the $A_H$ value towards their field using the N08 extinction
law fit to the RGC centroid in the OGLE-III CMD and the
$R_{JKVI}$ value from Table 3 of \citet{nat13} for their field.
The $A_H$ value we derived here is $A_H  = 0.122$, which is different from the value of $A_H = 0.265$
used by \citet{zoc03}. Therefore, we convert their extinction corrected $H$-band luminosity function
to a luminosity function with our preferred extinction model by adding $\Delta A_H = 0.265 - 0.122 = 0.142$
to their extinction corrected magnitudes, and then add the extinction appropriate for our
field, $A_H = 0.19$.

 We assume that stars can be resolved only if they are separated from the source by 
$\geq 0.8\,{\rm FWHM} = 148\,$mas. Under this assumption, the expected number of 
ambient stars within the circle is
derived by multiplying the area of this unresolvable region by the total number density derived above.
We determine the number of stars following the Poisson distribution with the 
mean value of the expected number of stars.
We use the corrected luminosity function to determine the magnitude  of each star.

\subsubsection{Source and lens companion flux priors}
We calculate the source and lens companion flux priors with the stellar binary distribution 
described  in \citet{kos17}. The binary distribution is based on the summary in a review 
paper \citep{duc13}, which provides distributions of the stellar multiplicity fraction, and
mass ratio and semi-major axis distributions.

For the flux of source companions, we calculate the source mass $M_{SC} = q_{SC} M_S$ and then 
convert that into  a source companion magnitude, $H_{SC}$, using 
a mass-luminosity relation. The mass ratio $q_{SC}$ is derived from the binary distribution.
We derive the source mass, $M_S,$ from the combination of $H_S$, $D_S$ and using the 
mass-luminosity relation.
Similarly, we calculate the lens companions magnitude, $H_{LC}$, from $M_{LC} = q_{LC} M_L$, 
where the lens mass $M_L$  comes from the same distribution that was used to obtain the lens
flux probability distribution.

We consider companions to the lens or source located in the same unresolvable regions in 
the vicinity of the source, just as in the case of ambient stars. Stellar companions have a
separation distribution that is much closer to logarithmic than the uniform distribution expected
for ambient stars. As a result, we must now exclude companions that are too close to the source 
and lens as well as companions
that are so widely separated that they will be resolved. Companions that are too close to the source
could be magnified themselves,  and companions that are too close to the lens could
serve as an additional lens star. Such a constraint
would have no effect on the ambient star probablity, because the probability of an
ambient star very close to the source or lens is much smaller than that of a  stellar companion.
Following \citet{bat14}, we adopt $\theta_E/4$ as the close limit for source companions 
and $w_{LC} < u_0$ as the close limit for lens companions, 
where $w_{LC} = 4q_{LC}/(s_{LC} - s_{LC}^{-1})^2$ \citep{chu05}
and $q_{LC}$ and $s_{LC}$ and are the stellar binary lens mass ratio and separation, respectively.
We take 0.8 FWHM as the maximum unresolvable radius.

We also consider triple and quadruple systems when estimating the effect of companions
to the source and lens, following \citet{kos17},  but we find
no significant difference from the case of only considering binary systems. 
We therefore do  not include triple and quadruple systems in this analysis, for simplicity.

\subsubsection{Excess flux prior}
\label{sec-excess_prior}
\begin{figure}
\centering
\epsscale{0.73}
\rotatebox{-90}{
\plotone{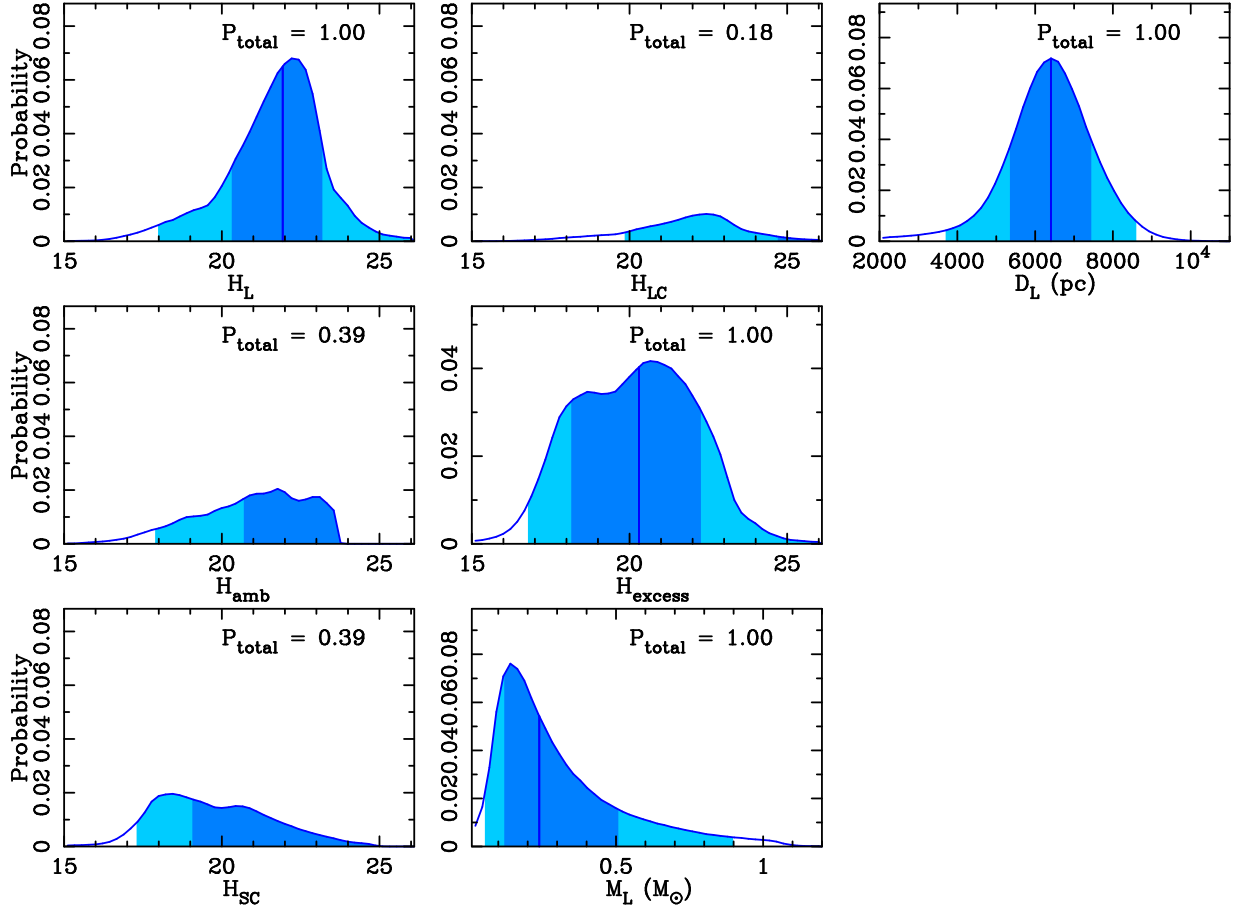}
}
\caption{The prior probability distributions using the assumptions in Table \ref{tab-assumption}
and light curve model constraints, as well as the seeing of the Keck AO image, but not
the target flux. We assume that 
the planet hosting probability does not depend on the stellar mass.
The borders between dark and light shaded regions indicate the $1 \sigma$ limits and the borders between 
light shaded and white regions indicate $2 \sigma$ limits.
The $P_{\rm total}$ value in each panel is the probability that the object exists. The panels with 
$P_{\rm total} < 1$ indicate the probability that the companion or ambient star actually exists,
and some of these do not have the
borders of the $1 \sigma$/$2 \sigma$ limit within the plotted region.}
\label{fig-prior}
\end{figure}

Figure \ref{fig-prior} shows the prior probability distributions we derived following the procedure
described above to calculate flux of each type of stars.
In addition to the magnitude of the 
four types of stars that might contribute to the excess flux, we show the prior distributions for 
the total excess flux, $H_{\rm excess}$, the lens mass, $M_L$, and the distance to the lens $D_L$.
Some of the panels in this figure have total probabilities $P_{\rm total} < 1$. This is because
many stars do not have binary companions and there is a large probability of no measurable
flux from an ambient star.
The $H_{\rm excess}$ prior indicates a high probability at the observed magnitude of 
$H_{\rm ex, obs} = 19.7 \pm 0.4$. The three panels for individual stars, $H_L$, $H_{\rm amb}$ and 
$H_{SC}$ show similar probabilities at the observed excess flux value. This indicates that it will
be difficult to claim that all of the excess comes from the lens itself.

\subsection{Posterior probability distributions}
\begin{figure}
\centering
\epsscale{0.73}
\rotatebox{-90}{
\plotone{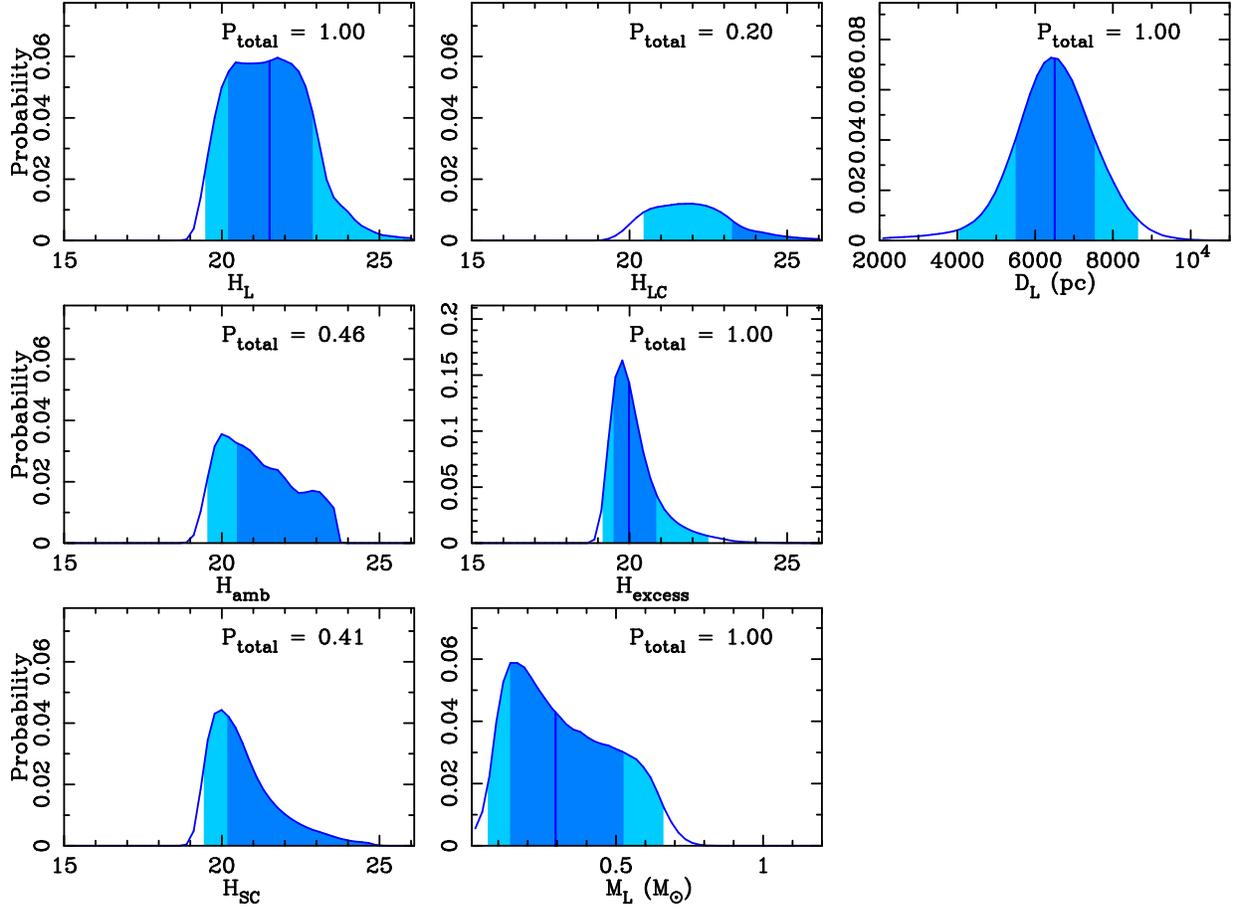}
}
\caption{The posterior probability distributions generated by extracting 
combinations which have consistent excess flux values with $H_{\rm ex, obs} = 19.7 \pm 0.4$ (in flux unit) from the prior probability distributions in Figure \ref{fig-prior}.}
\label{fig-post}
\end{figure}

We generate the posterior probability distributions shown in Figure \ref{fig-post} by
extracting combinations of parameters which have values of $H_{\rm excess}$ consistent with the
measured value of $H_{\rm ex, obs} = 19.7 \pm 0.4$ using a Gaussian distribution in fluxes (not magnitudes).
The probability that $H_L \leq 20$ is almost same as the probability for $H_{\rm SC}\leq 20$ and slightly
higher, but competitive with the probability that $H_{\rm amb} \leq 20$, which results in very loose 
constraints on $H_L$ and $M_L$.
This result is consistent with our expectation as discussed in Section~\ref{sec-excess_prior}.

The third to sixth columns of Table \ref{tab-prop} shows the median, the 1~$\sigma$ error bars, 
and the 2~$\sigma$ range for $H_L$, $M_L$ and $D_L$ for both the prior and posterior distributions.
This same table also shows the values of the planet mass $M_{\rm p}$, the projected 
separation $a_{\perp}$ and the three-dimensional star-planet separation $a_{3d}$ 
calculated from the probability distributions, where $a_{3d}$ is statistically estimated assuming a uniform
orientation for the detected planets. 
In the bottom three rows, we present the probabilities that the fraction of the excess flux due to the
lens, $f_L$ is larger than 0.1, 0.5 and 0.9, which correspond to 
magnitude difference between the lens and the total flux excess of 2.5 mag, 0.75 mag and 0.11 mag, respectively.

The posterior distributions for the lens system properties are remarkably similar to the prior distributions. 
When we compare the $1 \sigma$ ranges of the prior and posterior distributions, we see 
that the lens system is most likely to be composed an M or K dwarf star host and a gas-giant planet.
However the prior and posterior distributions differ from each other when we consider the $2 \sigma$ ranges
and the tails of the distributions. The possibility of a G dwarf host star is ruled out by the posterior distribution
while the host star can be a G dwarf according to  the prior distribution.
This implies that the host star is likely to be an M or K dwarf.

\subsection{Comparison of different planetary host priors}

One assumption that we have made implicitly is that the properties of the 
lens star do not depend on the fact that we have detected a planet orbiting the star.
This assumption could be false. Perhaps more massive stars are more likely to host
planets of the measured mass ratio, or perhaps disk stars are more likely to host 
planets than bulge stars. The microlensing method can be used to address these 
questions, but we must be careful not to assume the answer to them.

We have assumed that this detection
of the planetary signal does not bias any other property of the lens star, such as its mass or
distance. If there was a strong dependence of the planet hosting probability at the measured
mass ratio of $9.3^{+0.2}_{-0.1}\times 10^{-3}$, then this implicit prior could lead to incorrect
conclusions. 
Some theoretical papers based on core-accretion \citep{lau04, ken08} and  analyses of
exoplanets found by radial velocities 
\citep{joh10} have argued that gas giants are less frequently orbiting low-mass stars,
however,  the difference disappears when the planets are classified by their mass ratio,
$q$, instead of their mass. Nevertheless, since the host mass dependence of the planet
hosting probability is not well measured, we investigate how our results depend on the choice
of this prior.

\begin{figure}
\centering
\epsscale{0.73}
\rotatebox{-90}{
\plotone{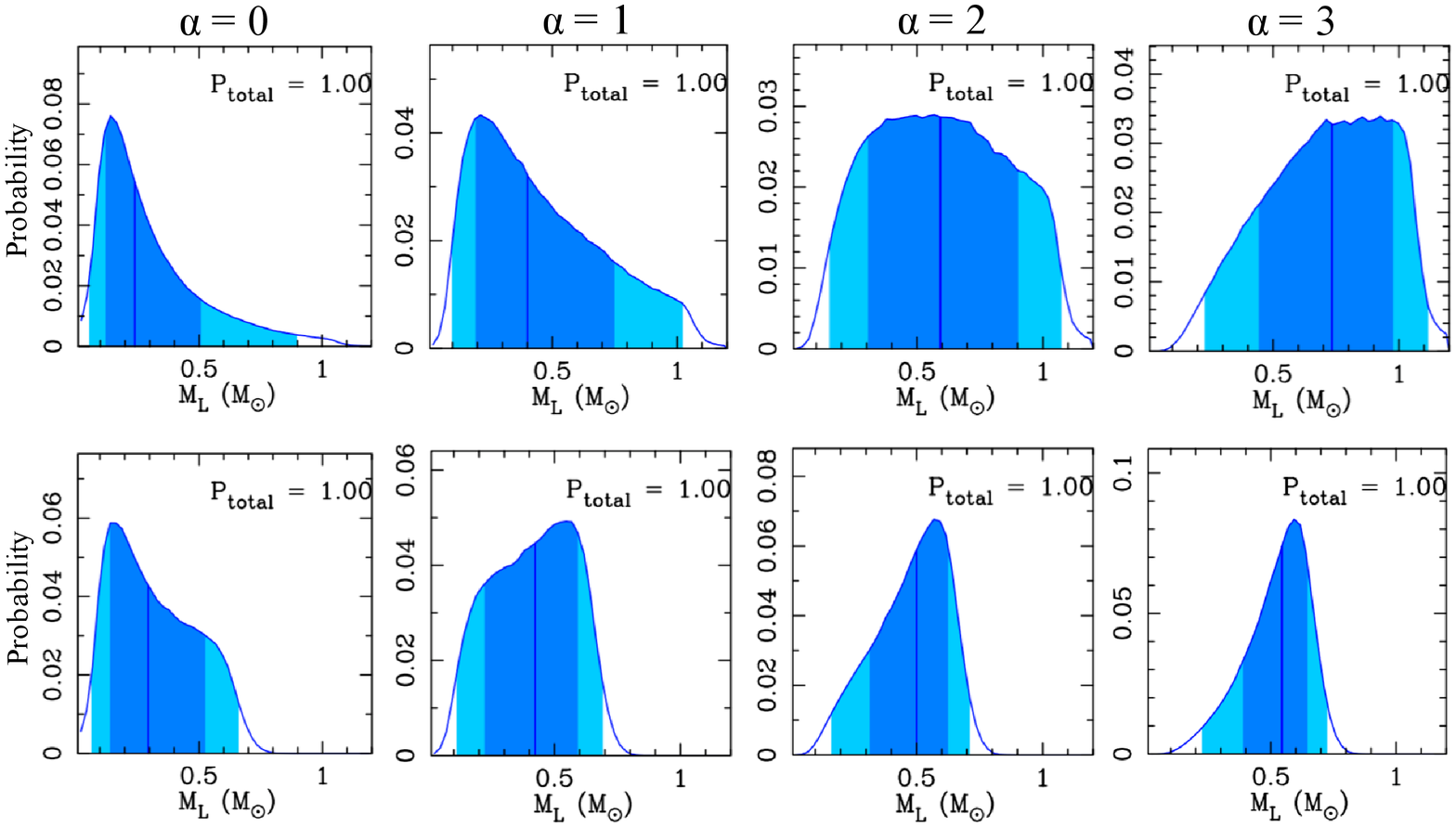}
}
\vspace{-2.5cm}
\caption{The prior (top) and posterior (bottom) probability distributions of the lens mass $M_L$ using different priors for the planet hosting probability, which is assumed to follow a power law,
$P_{\rm host} \propto M^{\alpha}$. The $\alpha = 0$ plots are repeated from 
Figures \ref{fig-prior} and \ref{fig-post}.
}
\label{fig-alpha}
\end{figure}

We consider a series of prior distributions where the
planet hosting probability follows a power law of the form
$P_{\rm host} \propto M^{\alpha}$, and we conduct a series of Bayesian analyses with $\alpha = 1$, 
$\alpha = 2$ and $\alpha = 3$ in addition to the calculation with $\alpha = 0$, presented above.
Figure \ref{fig-alpha} shows both the prior and posterior probability distributions for the lens mass, $M_L$, with these
different values of $\alpha$. The lens property values for each posterior distribution are shown in 
Table \ref{tab-prop}. The median of expected lens flux  approaches the measured excess flux as 
$\alpha$ increases (i.e., the power law becomes steeper), and consequently the median of the lens mass 
also increases and the parameter uncertainties decrease.
Thus, larger $\alpha$ values imply that more of the excess flux is likely to come from the lens. 
Nevertheless, our basic conclusion that the host is a M or K-dwarf hosting a gas giant planet
remains for all of the $1 \leq \alpha \leq 3$ priors.

\section{Discussion and Conclusion} \label{sec-disc}
We have analyzed the planetary microlensing event MOA-2016-BLG-227 which was discovered next to 
the field observed by the microlensing campaign (Campaign 9) of the {\it K2} Mission.
The event and planetary signal were discovered by the  MOA collaboration and a significant portion of
the planet signal was covered by the data from the Wise, UKIRT, CFHT and VST surveys, which
observed the event as part of the {\it K2}C9 program. 
Analysis of these data yields a unique microlensing light curve solution with
a relatively large planetary mass-ratio of $q = 9.28^{+0.20}_{-0.11} \times 10^{-3}$.
We considered several different extinction laws and decided that the N08 \citep{nis08} law
was the best fit to our data, although our results would not change significantly with a different law. 
With this extinction law, we derive an angular Einstein radius of $\theta_E = 0.227^{+0.006}_{-0.009}\,$mas,
which yields the mass-distance relation given in Equation~\ref{eq-m_thetaE}.
We detected excess flux at the location of the source in a Keck AO image, and we 
performed a Bayesian analysis to estimate the relative probability of different sources of this
excess flux, such as the lens, an ambient star, or a companion to the host or source.
Our analysis excludes the possibility that the host star is a G-dwarf, leading us to  a conclusion
that the planet MOA-2016-BLG-227Lb is a super-Jupiter mass planet orbiting
an M or K-dwarf star likely located in the Galactic bulge. Such systems are predicted to be rare
by the core accretion theory of planet formation.
It is also thought that such a planet orbiting a white dwarf host at $a_{3d} \sim 2$ AU is 
unlikely \citep{bat11}.

 If the planet frequency does not depend on the host star mass or
distance, our Bayesian analysis indicates the system consists of a host star with mass of 
$M_L = 0.29^{+0.23}_{-0.15} M_{\odot}$ orbited by a planet with mass of
$M_{\rm p} =2.8^{+2.2}_{-1.5} M_{\rm Jup}$ with a three-dimensional star-planet 
separation of $a_{3d} = 1.67^{+0.94}_{-0.35}$ AU. The system is located at $D_L = 6.5 \pm 1.0\,$kpc 
from the Sun. We also considered different priors for the planet hosting probability as a 
function of  host star mass. We  consider planet hosting prior probabilities that scale as 
$P_{\rm host} \propto M^{\alpha}$ with $\alpha = 1, 2, 3$, in 
addition to the $\alpha = 0$ prior that we use for our main results.  As $\alpha$ increases,
the median value of the lens mass also increases and the probability for the lens to be 
responsible for the excess $H$-band flux increases, as well.
\citet{joh10} found a linear (i.e., $\alpha = 1$) relationship between host mass and planet occurrence 
from 0.5 $M_{\odot}$ to 2.0 $M_{\odot}$ for giant planets within $\sim 2$AU around host stars
discovered by the radial velocity (RV) method. However, this analysis used a fixed minimum mass instead
of a fixed mass ratio, and it does not appear that \citet{joh10} did a detailed calculation of their
detection efficiencies. Another result using RV planet data by \citet{mon14} gives 
$\alpha = 0.8^{+1.1}_{-0.9}$, using a sample more similar to the microlensing planets, i.e., 
gas giants orbiting at $0 < a < 20$ AU around M-dwarf stars. However, our basic conclusion
that the MOA-2016-BLG-227L host star is an M or K-dwarf with a gas-giant planet 
located in the Galactic bulge would not change with a different $\alpha$ value, as
indicated in Figure~\ref{fig-alpha} and Table~\ref{tab-prop}.

The probability that more than 90\% of the excess flux seen in the Keck AO images comes from the 
lens is still 24.0\% even assuming $\alpha = 0$. This is significant enough that we cannot 
ignore the possibility that most of the excess flux comes from the lens star.
One approach for obtaining further constraints is to get the color of the excess flux. If the excess 
flux is not from the lens, the derived lens mass and distance with $H_{\rm excess}$ may be 
inconsistent with the value derived using the excess flux in a different pass band, if we assume that 
all of the excess flux comes from the lens. However, such a measurement could also yield
ambiguous results. Another, more definitive, approach is to observe this event in the future 
when we can expect to detect the lens-source
separation through precise PSF modeling with high resolution space-based data \citep{bennett07,ben15} 
or direct resolution with AO imaging \citep{bat15}.
The lens-source relative proper motion value of $\mu_{\rm rel} = 4.88^{+0.14}_{-0.17}\,$mas/yr 
indicates that we can expect to be able to resolve the lens, if it provides a large fraction of the excess flux
in $\sim 2022$ using {\it HST} \citep{bha17} and in 2026 using Keck AO \citep{bat15}.
Observations by the James Webb Space Telescope \citep{gar06}, the Giant Magellan Telescope \citep{joh12}, the Thirty
Meter Telescope \citep{nel08} and the Extremely Large Telescope \citep{gil07} could detect the lens-source relative
proper motion much sooner. 
If the separation of the excess flux from the source is different from the prediction of the microlensing
model in these future high angular resolution observations, it would indicates that the lens is not the 
main cause of the excess flux, implying a lower mass planetary host star.

Work by N.K. is supported by JSPS KAKENHI Grant Number JP15J01676. 
The MOA project is supported by grants JSPS25103508 and 23340064.
D.P.B., A.B., and D.S.\  were supported by NASA through grants NNX13AF64G, NNX16AC71G, 
and NNX16AN59G. 
Work by Y.S. and C.B.H. was supported by an appointment to the NASA Postdoctoral Program at the Jet
Propulsion Laboratory, California Institute of Technology, administered by Universities Space Research Association
through a contract with NASA. 
N.J.R. is a Royal Society of New Zealand Rutherford Discovery Fellow.
Work by C.R. was supported by an appointment to the NASA Postdoctoral Program at the Goddard Space Flight Center, administered by USRA through a contract with NASA.
Work by S.C.N. was supported by NExScI.
A.C. acknowledges financial support from Universit\'e
Pierre et Marie Curie under grants \'Emergence@Sorbonne Universit\'es 2016
and \'Emergence-UPMC 2012.
This work was supported by a NASA Keck PI Data Award, administered by the NASA Exoplanet
Science Institute. Data presented herein were obtained at the W. M. Keck Observatory from
telescope time allocated to the National Aeronautics and Space Administration through the
agency's scientific partnership with the California Institute of Technology and the University of
California. The Observatory was made possible by the generous financial support of the W. M.
Keck Foundation.
The authors wish to recognize and acknowledge the very significant cultural role and reverence that the summit of Mauna Kea has always had within the indigenous Hawaiian community.  We are most fortunate to have the opportunity to conduct observations from this mountain.
The United Kingdom Infrared Telescope (UKIRT) is supported by NASA and
operated under an agreement among the University of Hawaii, the University
of Arizona, and Lockheed Martin Advanced Technology Center; operations are
enabled through the cooperation of the Joint Astronomy Centre of the Science
and Technology Facilities Council of the U.K.
We acknowledge the support from NASA HQ for the UKIRT observations in connection with $K2$C9.
This research uses data obtained through the Telescope Access Program (TAP), which has been funded by the National Astronomical Observatories of China, the Chinese Academy of Sciences (the Strategic Priority Research Program ``The Emergence of Cosmological Structures'' Grant No. XDB09000000), and the Special Fund for Astronomy from the Ministry of Finance. 
This work is partly based on observations obtained with MegaPrime/MegaCam, a joint project of CFHT and CEA/DAPNIA, at the Canada-France-Hawaii Telescope (CFHT) which is operated by the National Research Council (NRC) of Canada, the Institut National des Science de l'Univers of the Centre National de la Recherche Scientifique (CNRS) of France, and the University of Hawaii.
This work was performed in part under contract with the California Institute of Technology (Caltech)/Jet Propulsion Laboratory (JPL) funded by NASA through the Sagan Fellowship Program executed by the NASA Exoplanet Science Institute. Work by MTP and BSG was supported by NASA grant NNX16AC62G.
This work was partly supported by the National Science Foundation of China (Grant No. 11333003, 11390372 to SM).
Based on observations made with ESO Telescopes at the La Silla Paranal
Observatory under programme ID 097.C-0261.

\appendix

\section{Comparison of Different Extinction Laws}\label{sec-complaw}
In Section \ref{sec-ext}, we obtained the observed extinction value,
$A_{I, {\rm obs}} = 0.98 \pm 0.07$, and color excess values of 
$E(V-I)_{\rm obs} = 0.82 \pm 0.04$, $E(V-H)_{\rm obs} = 1.67 \pm 0.11$ and $E(I-H)_{\rm obs} =0.81 \pm 0.07$.
Then, we fit these values to the extinction laws of \citet{car89}, \citet{nis09} and \citet{nis08} 
separately and compared the results. This was motivated by the fact that \citet{nat16} reported 
a clear difference of their extinction law towards the Galactic bulge from the standard law of \citet{car89}.
Hereafter, we refer to these papers as C89, N09 and N08, respectively.
Note that the four observed extinction parameters (1 extinction and 3 color excess) are not independent.
They can be derived from the three independent extinction values:
$A_{I, {\rm obs}}$, $A_{V, {\rm obs}}$ and $A_{H, {\rm obs}}$.

The C89 law is given by equations (1) - (3b) in their paper, and $A_V$ and $R_V$ serve as the parameters
of their model.

Unlike C89, N09 does not provide a complete extinction model. They provide only 
ratios of extinctions for wavelengths longer than the $J$-band. So,
we need additional information relating $A_V$ or $A_I$ and $A_J$, $A_H$ or $A_K$ in order
to calculate the values that we need for this paper: $A_I$, $A_V$ and $A_H$.
Therefore we used the $R_{JKVI} \equiv E(J-K_s)/E(V-I)$ values from \citet{nat13} 
in addition to the N09 extinction law.
The $R_{JKVI}$ value at the nearest grid point to the MOA-2016-BLG-227 event in 
Table 3 of \citet{nat13} is 0.3089. However the quality flag for this value is 1,
which indicates an unreliable measurement, so we use a conservative uncertainty of
$R_{JKVI} = 0.31 \pm 0.03$.
We adjust $A_I$ and $E(V-I)$ to minimize the $\chi^2$ value between the observed
$A_{I,\rm obs}$, $E(V-I)_{\rm obs}$, $E(V-H)_{\rm obs}$, and $E(I-H)_{\rm obs}$ values and 
those values derived using the ratio $A_{H, {\rm 2MASS}}/E(J - K_s)_{\rm 2MASS}= 0.89$ from 
N09, in conjunction with the $R_{JKVI}$ value from \citet{nat13}.
We explicitly use the $_{\rm 2MASS}$
subscript because N09 provides their result
also in the IRSF/SIRIUS photometric system \citep{nag99, nag03}.
Note that we calculate this $A_H/E(J-K_s)$ value using their result
for the field S+ ($0^{\circ} < l < 3^{\circ}, -1^{\circ} < b < 0^{\circ}$), which is nearest of their
fields to the MOA-2016-BLG-227 event position.

N08 also provide the ratio of extinctions towards the Galactic bulge 
($l \sim 0^{\circ}, b \sim -2^{\circ}$). They find $A_J/A_V = 0.183 \pm 0.015$, 
$A_H/A_V = 0.103 \pm 0.008$ and $A_{K_s}/A_V = 0.064 \pm 0.005$.
(These values are slightly different from the original values given by N08 
because the values used in N08 were in the OGLE II and IRSF/SIRIUS photometric systems, so
we converted them into the standard systems that we use here.)
These values are well fit by a single power law, $A_{\lambda}/A_V \propto \lambda^{-2}$.
Nevertheless, we use the ratios themselves, instead of the
single power law, because N08 does not test that their power law accurately reproduces
$A_I/A_V$, which we have now. As in the case of N09, we keep these ratios fixed, and adjust 
$A_I$ and $E(V-I)$ to minimize the $\chi^2$ between these relations and the observed
$A_{I,\rm obs}$, $E(V-I)_{\rm obs}$, $E(V-H)_{\rm obs}$, and $E(I-H)_{\rm obs}$ values.
Notice that N08 had $V$-band data and it was not necessary to use the $R_{JKVI}$ as a constraint.
Therefore we used the $R_{JKVI}$ value as the additional observed data instead here in 
addition to $A_{I, {\rm obs}}$, $A_{V, {\rm obs}}$ and $A_{H, {\rm obs}}$ to increase  number of 
degrees of freedom (dof).

Table \ref{tab-ext} shows the results of fitting our extinction measurements to these 
three different extinction laws. This table also shows the angular source radius calculated from the 
extinction--corrected source magnitudes and colors using formulae from the analysis of \citet{boy14}.
We determine $\theta_{*,IH}$ using Equations (1)-(2) and Table 1 of \citet{boy14}, but the other
relations were provided by private communications from Boyajian with a special
analysis restricted to stellar colors that are relevant for the Galactic bulge sources
observed in microlensing events.
We use Equation (4) of \citet{fuk15} to determine $\theta_{*, VI}$, and
we use  Equation (4) of \citet{ben15} to determine $\theta_{*,VH}$.
Those formulae are
\begin{align}
\log ~ [2 \theta_{*, VI}/(1 {\rm mas})] &= 0.5014 + 0.4197 (V-I)_{S,0} - 0.2 I_{S,0},  \\
\log ~ [2 \theta_{*, VH}/(1 {\rm mas})] &= 0.5367 + 0.0727 (V-H)_{S,0} - 0.2 H_{S,0},  \\
\log ~ [2 \theta_{*, IH}/(1 {\rm mas})] &= 0.5303 + 0.3660 (I-H)_{S,0} - 0.2 I_{S,0}.  
\end{align}

If we compare the $\chi^2$ value for each model fit in Table~\ref{tab-ext}, we see that the 
$\chi^2/{\rm dof}$ for the N09 and N08 laws are smaller than the value from the C89 extinction law, 
although the C89 is not disfavored by a statistically significant amount. 
(The $p$-value of $\chi^2 = 2.39$ for ${\rm dof} = 1$ is still $\sim$0.12.)
Note that a contribution of $0.96$ to the total value of $\chi^2 = 1.19$ arises from fitting 
the $R_{JKVI}$ value to the N08 extinction law. 
So, the remaining contribution of 0.23 to $\chi^2$ arises from fitting the N08 model to our measurements of the RGC centroids.
This indicates that the extinction law of N08 agrees with our measurement of the red clump centroids very well, 
but not quite so well with the $R_{JKVI}$ value, which comes from \citet{nat13}.

From the point of view of consistency between the three $\theta_*$ values, the standard deviation of the three values (SD$_{\theta_*}$ in the table) is smallest 
using the N08 extinction laws. The N08 extinction law also yields the smallest error bars 
for $A_H$ and $\theta_{*, VH}$.

Based on this analysis, we have decided to use the results from the N08 extinction laws
in our analysis.
We use $\theta_{*,VH}$ for the final angular source radius which is $\theta_* = 0.68 \pm 0.02 ~ \mu$as.
We show the source magnitudes and colors corrected for extinction using the N08 extinction laws in 
Table \ref{tab-RCG}.

\clearpage



\clearpage

\clearpage

\clearpage

\clearpage

\clearpage

\clearpage

\clearpage



\catcode`?=\active \def?{\phantom{0}}
 \tblcaption{The data and the parameters for the modeling.}
\vspace{-0.4cm}
\begin{center}
\begin{threeparttable}
 \begin{tabular}{lrrrrr}\hline\hline
Dataset    & Number of data & $k$   & $e_{\rm min}$ & $u_{\lambda}$ \\\hline
MOA-Red\tnote{a}  & 1804    & 0.938 & 0             & 0.5585\\
MOA $V$               & 60         & 1.224 & 0             & 0.6822\\
Wise Astrodon\tnote{b}  & 44  & 1.267 & 0             & 0.6015\tnote{b}\\
UKIRT $H$    & 127               & 1.072 & 0.015       & 0.3170\\
CFHT $i$    & 77                   & 1.673\tnote{c} & 0.003      & 0.5360\\
CFHT $r$    & 77                   & 3.028\tnote{c} & 0             & 0.6257\\
CFHT $g$    & 78                   & 2.105\tnote{c} & 0             & 0.7565\\
VST-71 $r$\tnote{d}  &  193  & 1.018 & 0            & 0.6257\\
VST-95 $r$\tnote{d}  &   97   & 1.080 & 0            & 0.6257\\\hline
    \end{tabular}
    \begin{tablenotes}
    \item{\bf Notes.} Parameters $k$ and $e_{\rm min}$ are used for the error normalization, and 
    $u_{\lambda}$ is the limb darkening coefficient. 
    \footnotesize
    \item[a] Approximately Cousins $R + I$.
    \item[b] This filter blocks $\lambda < 500\,$nm, and we use the limb darkening coefficient $u_R$
    to describe limb darkening in this filter.
    \item[c] The CFHT error estimates were underestimated by a constant factor of 1.54, resulting in larger values of the $k$ parameters.
    \item[d] These use the same SDSS $r$ filter, but different detectors, numbers 71 and 95, respectively.
    \end{tablenotes}
    \end{threeparttable}
    \end{center}
    \label{tab-data}
  \hfill

\catcode`?=\active \def?{\phantom{0}}
 \tblcaption{The parameters for the best-fit binary lens model.}
\vspace{-0.4cm}
\begin{center}
\begin{threeparttable}
   \begin{tabular}{ccccccccccc}\hline\hline
Parameter & Unit  & Value\\\hline
$t_0$ & HJD - 2450000  & $7517.5078 ^{+0.007}_{-0.006}$ \\
$t_E$ & days         &  $17.03 ^{+0.08}_{-0.20} $ \\
$u_0$ & $10^{-2}$ & $ -8.33^{+0.08}_{-0.16}$ \\
$q$    &  $10^{-3}$ & $ 9.28^{+0.20}_{-0.11}$ \\
$s$    &                     & $ 0.9312^{+0.0004}_{-0.0009}$ \\
$\alpha$  & rad & $ 2.509^{+0.003}_{-0.004}$ \\
$\rho$    &  $10^{-3}$ &  $ 3.01^{+0.09}_{-0.05}$ \\
$\chi^2$ &  & 2538.9\\
${\rm dof}$  & & 2538\\\hline
    \end{tabular}
    \begin{tablenotes}
    \item{\bf Notes.} Superscripts and subscript indicates the the 84th and 16th percentile from the best-fit values, respectively. 
   \end{tablenotes}
    \end{threeparttable}
    \end{center}
    \label{tab-models}
  \hfill

\clearpage

\small
\catcode`?=\active \def?{\phantom{0}}
 \tblcaption{The source and RGC magnitude and colors.}
\vspace{-0.4cm}
\begin{center}
\begin{threeparttable}
   \begin{tabular}{lcccccccccccc}\hline\hline
                                                         &  $I$                         & $V - I$                 & $V -H$               &  $I - H$\\\hline
RGC (measured from CMD)            &  $15.33 \pm 0.05$ &  $1.88 \pm 0.02$ & $4.03 \pm 0.06$ &  $2.11 \pm 0.03$\\
RGC (intrinsic)                                 &  $14.36 \pm 0.05$ &  $1.06 \pm 0.03$ &  $2.36 \pm 0.09$ &  $1.30 \pm 0.06$\\
Source (measured from lightcurve) &  $19.54 \pm 0.02$ &  $1.60 \pm 0.03$ & $3.33 \pm 0.03$ &  $1.73 \pm 0.02$\\
Source (intrinsic) \tnote{a}               &  $18.54 \pm 0.09$ &  $0.78 \pm 0.06$ &  $1.70 \pm 0.11$ &  $0.92 \pm 0.08$\\\hline
    \end{tabular}
      \begin{tablenotes}
    \footnotesize
    \item[a] Extinction corrected magnitudes using the \citet{nis08} extinction model from Table \ref{tab-ext}
       \end{tablenotes}
      \end{threeparttable}
    \end{center}
    \label{tab-RCG}
  \hfill

{\tabcolsep = 1.2mm
\normalsize
\catcode`?=\active \def?{\phantom{0}}
 \tblcaption{Assumptions and undetectable limits used for the prior probability distributions}
\vspace{-0.4cm}
\begin{center}
\begin{threeparttable}
    \begin{tabular}{ccccl}\hline\hline
Priors of                & Assumption              & Closer limit & Wider limit & Used observed values\\\hline
$H_L$                           & Galactic model         &       $-$      &       $-$       &  $t_{\rm E}$, $\theta_{\rm E}$ \\
??$H_{\rm amb}$         & Luminosity function  &       $-$      & 0.8 FWHM & FWHM, Number density\\
?$H_{SC}$                    & Binary distribution \tnote{a} & $\theta_E$/4 & 0.8 FWHM & FWHM, $\theta_E$, $H_S$\\
?$H_{LC}$                     & Binary distribution \tnote{a} & $w_{LC} \tnote{b} < u_0$ & 0.8 FWHM & FWHM, $\theta_E$, $u_0$, $M_L$\tnote{c}\\\hline
    \end{tabular}
     \begin{tablenotes}
     \footnotesize
    \item[a] The binary distribution used by \citet{kos17}, based on \citet{duc13}.
    \item[b] The caustic size created by the hypothetical companion to the lens, $w_{LC} = 4q_{LC}/(s_{LC} - s_{LC}^{-1})^2$.
    \item[c] The $M_L$ value extracted from the prior probability distributions to calculate the $H_L$ value.
    \end{tablenotes}
    \end{threeparttable}
    \end{center}
\label{tab-assumption}
}

\footnotesize
\catcode`?=\active \def?{\phantom{0}}
 \tblcaption{Lens properties calculated from the prior and posterior probability distributions. }
\vspace{-0.4cm}
\begin{center}
\begin{threeparttable}
    \begin{tabular}{ccccccccc}\hline\hline 
Parameters        & Unit                  & Prior                                & $2 \sigma$ range & Posterior                      & $2 \sigma$ range & Posterior                         & Posterior                           & Posterior       \\
                           &                          &   $\alpha = 0$                 &                              &    $\alpha = 0$             &                              &    $\alpha = 1$                 &  $\alpha =2$                     &     $\alpha =3$           \\\hline
$H_L$                & mag                  & $21.9^{+1.3}_{-1.6}$    & 18.0-28.7             & $21.5^{+1.4}_{-1.3}$      &  19.5-27.5              & $20.8^{+1.3}_{-0.9}$      &  $20.4^{+1.1}_{-0.7}$    & $20.2^{+0.8}_{-0.6}$ \\
$M_L$               & $M_{\odot}$     & $0.24^{+0.27}_{-0.12}$ & 0.06-0.90             & $0.29^{+0.23}_{-0.15}$  &  0.07-0.66              & $0.42^{+0.17}_{-0.20}$  &  $0.50^{+0.13}_{-0.18}$ &  $0.54^{+0.10}_{-0.15}$\\
$M_{\rm p}$       & ?$M_{\rm Jup}$ & $2.3^{+2.6}_{-1.2}$       &  0.5-8.8                & $2.8^{+2.2}_{-1.5}$        &  0.6-6.4               & $4.1^{+1.7}_{-1.9}$        &  $4.8^{+1.2}_{-1.8}$       &  $5.3^{+1.0}_{-1.5}$    \\
$D_L$                & kpc                   & $6.4 \pm 1.0$               &   3.5-8.5               & $6.5 \pm 1.0$                  &  3.9-8.6                 & $6.8^{+1.0}_{-0.9}$        &  $6.9^{+1.0}_{-0.9}$        &  $7.1^{+1.0}_{-0.9}$  \\
$a_{\perp}$        & AU                    & $1.37 \pm 0.23$           &  0.76-1.84            & $1.39 \pm 0.22$              &  0.84-1.86             & $1.45^{+0.22}_{-0.20}$  &  $1.49^{+0.22}_{-0.20}$  &  $1.51^{+0.22}_{-0.20}$\\
?$a_{3d}$   & AU                    & $1.64^{+0.93}_{-0.36}$ &  0.89-6.49            & $1.67^{+0.94}_{-0.35}$  & 0.97-6.62            & $1.74^{+0.99}_{-0.35}$  & $1.79^{+1.02}_{-0.35}$  &  $1.82^{+1.03}_{-0.36}$\\\hline
$P(f_L > 0.1)$\tnote{a} &  \%        &  72.2                               &              -                 &         78.1                    &            -                  &             90.8                       &            96.5                   &            98.7                \\
$P(f_L > 0.5)$\tnote{a} &  \%        &  48.0                               &              -                 &         41.4                     &           -                   &            56.7                       &            69.6                    &           78.4                 \\
$P(f_L > 0.9)$\tnote{a} &  \%        &  33.8                               &              -                 &         24.0                     &           -                   &            29.9                       &            38.0                    &           44.5               \\\hline
    \end{tabular}
     \begin{tablenotes}
     \small
     \item{\bf Notes.}  The values of posterior probability distributions are shown also for different $\alpha$ values, the slope of the probability of hosting planets $P_{\rm host} \propto M^{\alpha}$.
     The values given in form of the median with the 1 $\sigma$ uncertainty. The 2 $\sigma$ range is given for $\alpha = 0$.
     \footnotesize
    \item[a]  The probabilities that the fraction of the lens flux to the excess flux,  $f_L \equiv F_L/F_{\rm excess}$, is larger than the indicated values. The fractions of 0.1, 0.5 and 0.9 correspond to 
                 the difference of magnitude, $H_L - H_{\rm excess} = -2.5~\log (F_L/F_{\rm excess})$, of 2.5 mag, 0.75 mag and 0.11 mag.
    \end{tablenotes}
    \end{threeparttable}
    \end{center}
\label{tab-prop}

\clearpage

{\tabcolsep = 1.2mm
\footnotesize
\catcode`?=\active \def?{\phantom{0}}
 \tblcaption{Comparison of the extinction and angular Einstein radius based on different extinction laws.}
\vspace{-0.4cm}
\begin{center}
\begin{threeparttable}
   \begin{tabular}{cccccccccccc}\hline\hline
Extinction law        &   None\tnote{a}   &   \citet{car89}                                                                            & \citet{nis09}                           &  \citet{nis08}  \\\hline
Relation                    &         -                  &   $\frac{A_\lambda}{A_V} = a(x) + \frac{b(x)}{R_V}$\tnote{b}  &  $\frac{A_H}{E(J-K_s)}$, $R_{JKVI}$\tnote{c} &   $\frac{A_J}{A_V}$, $\frac{A_H}{A_V}$, $\frac{A_{K_s}}{A_V}$\\\hline
$A_V$                    &  $1.80 \pm 0.08$ &  $1.87 \pm 0.12$                                                                    &  $1.83 \pm 0.12$                    & ${\bf 1.82 \pm 0.12}$ \\
$A_I$                     &  $0.98 \pm 0.07$ &  $1.04 \pm 0.10$                                                                    &  $1.01 \pm 0.08$                   & ${\bf 1.00 \pm 0.08}$ \\
$A_H$                    &  $0.17 \pm 0.10$ &  $0.30 \pm 0.04$                                                                    &  $0.23 \pm 0.04$                   & ${\bf 0.19 \pm 0.02}$ \\
$E(V-I)$                  &  $0.82 \pm 0.04$ &  $0.83 \pm 0.05$                                                                    &  $0.82 \pm 0.05$                   & ${\bf 0.82 \pm 0.05}$ \\
$E(V-H)$                &  $1.67 \pm 0.11$ &  $1.57 \pm 0.10$                                                                    &  $1.60 \pm 0.11$                   & ${\bf 1.63  \pm 0.11}$ \\
$E(I-H)$                 &  $0.81 \pm 0.07$ &  $0.73 \pm 0.06$                                                                    &  $0.78 \pm 0.09$                   & ${\bf 0.81 \pm 0.08}$ \\
$\chi^2 / {\rm dof}$ \tnote{d}  &  -                          & 2.39/1                                                                                     &  0.56/1                                    & 1.19/2\tnote{e} \\\hline
$\theta_{*, VI} $  ($\mu$as)  &  $0.65 \pm 0.04$ &  $0.67 \pm 0.06$                                                     &  $0.66 \pm 0.05$                   & $0.66 \pm 0.05$ \\
$\theta_{*, VH} $ ($\mu$as) &  $0.67 \pm 0.04$ &  $0.73 \pm 0.02$                                                     &  $0.70 \pm 0.02$                   & ${\bf 0.68 \pm 0.02}$ \\
$\theta_{*, IH} $  ($\mu$as)  &  $0.71 \pm 0.07$ &  $0.79 \pm 0.06$                                                    &  $0.74 \pm 0.06$                   & $0.72 \pm 0.06$ \\
SD$_{\theta_*}$\tnote{f}      &  0.035                  & 0.061                                                                      &  0.042                                    & 0.030             \\\hline
$\theta_{\rm E}$ (mas) \tnote{g}    &        -                              &               -                  &                   -                                                                      &     ${\bf 0.227^{+0.006}_{-0.009}}$ \\
$\mu_{\rm rel}$ (mas/yr) \tnote{g} &         -                                &              -                   &                     -                                                                    &     ${\bf 4.88^{+0.14}_{-0.17}}$ \\\hline
    \end{tabular}
    \begin{tablenotes}
    \small
     \item{\bf Notes.} The values in boldface are used as final values.
    \footnotesize
    \item[a] Result without using an extinction law. The $A_I$, $E(V-I)$, $E(V-H)$ and $E(I-H)$ values are determined directly from the data.
    \item[b] Equation (1) of C89, see the paper for the detailed model.
    \item[c] The $R_{JKVI}$ value comes from Table 3 of \citet{nat13}.
    \item[d] When calculating the total $\chi^2$, we multiply each of the contributions from $E(V-I)$, $E(V-H)$ and $E(I-H)$ by 2/3, because these values are not independent.
    \item[e] The ${\rm dof} = 2$ is because we used the $R_{JKVI}$ value from \citet{nat13} as an observed data point.
    \item[f] Standard deviation of the three $\theta_*$ values.
    \item[g] Calculations conducted only for the adopted $\theta_*$ value ($\theta_{*, VH}$ with N08).
   \end{tablenotes}
    \end{threeparttable}
    \end{center}
    \label{tab-ext}
  \hfill
}

\end{document}